\documentclass[twocolumn,english,aps,prx,longbibliography,superscriptaddress]{revtex4-2}
\usepackage[T1]{fontenc}
\usepackage{color}
\usepackage{xcolor}
\usepackage{mathtools}
\usepackage{amsmath,amssymb,mathrsfs}
\usepackage{graphicx}
\usepackage{natbib}
\usepackage{braket}
\usepackage[normalem]{ulem}
\usepackage{float}
\usepackage[colorlinks=true,urlcolor=blue,citecolor=blue,linkcolor=blue]{hyperref}

\makeatother

\usepackage{babel}
\newcommand\scalemath[2]{\scalebox{#1}{\mbox{\ensuremath{\displaystyle #2}}}}

\begin{document}
\title{Robust Magnetometry with Single NV Centers via Two-step Optimization}
\author{Nimba Oshnik}
\affiliation{Technische Universit{\"a}t Kaiserslautern, Department of physics, Erwin Schr{\"o}dinger Strasse, D-67663 Kaiserslautern, Germany}
\author{Phila Rembold}
\affiliation{Dipartimento di Fisica e Astronomia ''G. Galilei'', Universit{\`a} degli Studi di Padova, I-35131 Padua, Italy}
\affiliation{Institute for Theoretical Physics, University of Cologne, D-50937 Cologne, Germany}
\affiliation{Istituto Nazionale di Fisica Nucleare (INFN), Sezione di Padova, I-35131 Padua, Italy}
\affiliation{Forschungszentrum J{\"u}lich GmbH, Peter Gr{\"u}nberg Institute -  Quantum Control (PGI-8), D-52425 J{\"u}lich, Germany}
\author{Tommaso Calarco}
 \affiliation{Forschungszentrum J{\"u}lich GmbH, Peter Gr{\"u}nberg Institute -  Quantum Control (PGI-8), D-52425 J{\"u}lich, Germany}
\affiliation{Institute for Theoretical Physics, University of Cologne, D-50937 Cologne, Germany}
\author{Simone Montangero}
 \affiliation{Dipartimento di Fisica e Astronomia ''G. Galilei'', Universit{\`a} degli Studi di Padova, I-35131 Padua, Italy}
\affiliation{Istituto Nazionale di Fisica Nucleare (INFN), Sezione di Padova, I-35131 Padua, Italy}
\affiliation{Padua Quantum Technologies Research Center, Universit{\`a} degli Studi di Padova, I-35131 Padova, Italy}
\author{Elke Neu}
\email{nruffing@rhrk.uni-kl.de}
\affiliation{Technische Universit{\"a}t Kaiserslautern, Department of physics, Erwin Schr{\"o}dinger Strasse, D-67663 Kaiserslautern, Germany}
\author{Matthias M. M{\"u}ller}
 \affiliation{Forschungszentrum J{\"u}lich GmbH, Peter Gr{\"u}nberg Institute -  Quantum Control (PGI-8), D-52425 J{\"u}lich, Germany}

\begin{abstract}
Shallow Nitrogen-Vacancy (NV) centers are promising candidates for high-precision sensing applications; these defects,  when positioned a few nanometers below the surface, provide an atomic-scale resolution along with substantial sensitivity. However, the dangling bonds and impurities on the diamond surface result in a complex environment which reduces the sensitivity and is unique to each shallow NV center. To avoid the environment's detrimental effect, we apply feedback-based quantum optimal control. We first show how a direct search can improve the initialization/readout process. In a second step, we optimize microwave pulses for pulsed Optically Detected Magnetic Resonance (ODMR) and Ramsey measurements. Throughout the sensitivity optimizations, we focus on robustness against errors in the control field amplitude. This feature not only protects the protocols' sensitivity from drifts but also enlarges the sensing volume. The resulting ODMR measurements produce sensitivities below 1$\mu$T\,Hz$^{-\frac{1}{2}}$ for an 83\% decrease in control power, increasing the robustness by approximately one third. The optimized Ramsey measurements produce sensitivities below 100\,nT\,Hz$^{-\frac{1}{2}}$ giving a two-fold sensitivity improvement. Being on par with typical sensitivities obtained via single NV magnetometry, the complementing robustness of the presented optimization strategy may provide an advantage for other NV-based applications. 
\end{abstract}

\maketitle
\section{\label{sec:1}Introduction}
Quantum sensing with NV~\footnote{In this manuscript, the term NV center denotes the negatively charged state of the nitrogen-vacancy center in diamond.} centers have evolved into a prominent branch of quantum technologies in the last two decades~\cite{Jelezko2006, Doherty2013, Degen2017, Barry2019, Rembold2020}. NV centers serve as a multipurpose sensor for detecting magnetic~\cite{Childress2006, Balasubramanian2008, Maze2008, Taylor2008} and electric fields~\cite{Dolde2011}, temperature~\cite{Acosta2010, Konzelmann2018}, and pressure~\cite{Doherty2014, Lesik1359}. Additionally, NV centers find applications as quantum memories~\cite{Fuchs2011}, quantum registers~\cite{Bradley2019}, and in other areas of emerging quantum technologies~\cite{Rembold2020, Wang2015,Ledbetter2012}. Rapid improvement in nano-fabrication methods~\cite{Balasubramanian2009, Degen2008, Bluvstei2019}, material science research~\cite{Osterkamp2019, Schreck2017}, as well as control methodologies~\cite{Degen2017, Rembold2020, Vandersypen2005, Glaser2015, Mueller2021} have led to a variety of NV-based quantum sensors with applications in the fields of life sciences~\cite{Schirhagl2014,Mohan2010}, and material studies~\cite{Nelz2020}.
\begin{figure*}
    \centering
    \includegraphics[width = 0.99\textwidth]{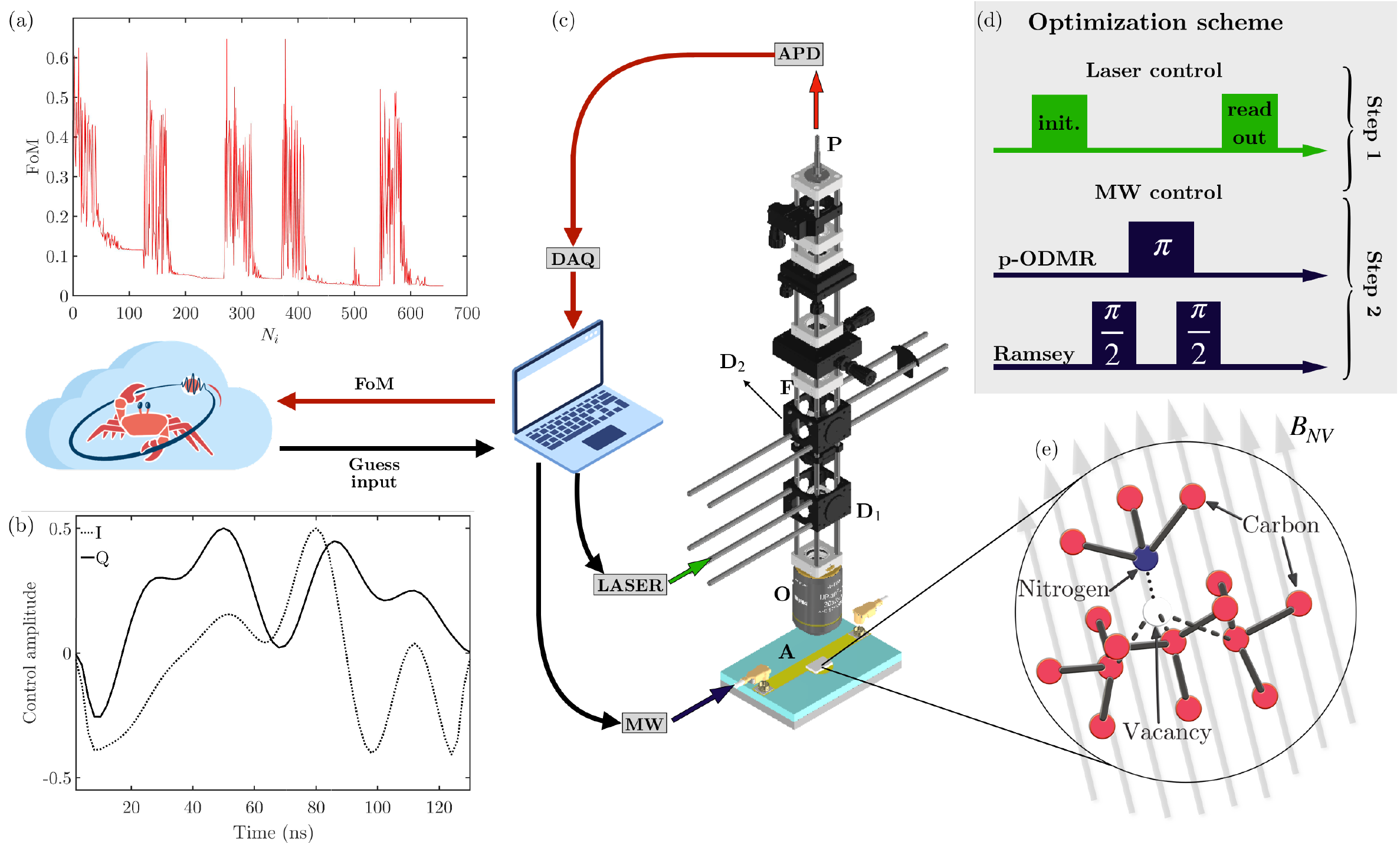}
    \caption{\label{fig:1}Schematic for the closed-loop optimization with single NV centers in diamond. The optimization algorithm suggests control pulses/parameters to the setup. The resulting Figure of Merit (FoM) is calculated from the output and passed back to the optimizer. This cycle repeats until the FoM converges. (a) An exemplary plot of the convergence of the FoM with the number of algorithm iterations $N_i$; the algorithm suggests different controls to find the global optimal solution. (b) The in-phase and quadrature components (I and Q) of a typical guess for a MW control pulse suggested by the algorithm. (c) The confocal setup used in combination with the RedCRAB optimization program; laser (green arrow) and MW (blue arrow) pulses are used to control the NV spin state. The fluorescence (red arrow) is collected with an optical fiber, $P$, connected to a single photon counter (APD), logged with a data acquisition device (DAQ), and further processed on the local control system to pass the FoM to the remote optimization server. (d) The two-step optimization strategy introduced in this work. In step 1 the laser based spin state initialization and readout processes are optimized. Step 2 creates robust MW control pulses for pulsed ODMR (p-ODMR) and Ramsey sensing sequences via QOC. (e) Lattice structure of the NV center. The NV quantization axis is shown as a dotted black line. The component of the external magnetic DC field along the NV quantization axis is denoted as $B_\text{NV}$ and quantified via the sensing methods. The confocal schematic in (c) is is drawn with parts adapted and modified from Ref.~\cite{thorlabs} and Ref.~\cite{grabcad} with permission under terms of reuse. For details on the setup see appendix~\ref{appa}.  }
\end{figure*}
NV centers exhibit optical spin-state polarization and spin-state dependent fluorescence~\cite{Childress2006, Jelezko2006, Balasubramanian2008}. Additionally, the NV spin-state can be manipulated with resonant microwave (MW) control fields.
Various sensing protocols are available that use MW-based unitary gates under the two-level approximation~\cite{Degen2017}. However, state-of-the-art NV-based quantum sensors do not perform on par with their theoretical potential. Because of the potential applications, further improvement of NV magnetometry is a flourishing and multidisciplinary research topic~\cite{Degen2017,Barry2019, Rembold2020}. While the NV centers particularly close to the surface may offer high nanoscale resolution~\cite{Okai2012,Romach2015}, they also exhibit especially short dephasing and decoherence times.

Likewise, limitations and errors related to the experimental setup, such as drift, finite bandwidth, and transfer functions, restrict the performance of these sensing methods. For example, to exploit the full potential of NV-based scanning probe applications, the MW antenna has to be brought close to the cantilever~\cite{Maletinsky2012, Appel2016, Zhou2017}, which can be experimentally challenging given the microscopic scale of the scanning devices. If the distance between antenna and cantilever is larger, it reduces the contrast and hence, the sensitivity of the setup. Additionally, applications with NV-based scanning probes~\cite{Schell2014, ariyaratne2018} that move with respect to the antenna experience variations in control power. The power variations, in turn, lead to a correspondingly worsened sensitivity. Similarly, applications with single NV centers~\cite{Lovchinsky2016, Bonato2016} or ensembles of NV centers~\cite{chipaux_magnetic_2015, Barry2019, Schlussel2018} in bulk diamond are subject to variation in control power depending on the distance from the MW antenna. In all these cases, robustness against control power variation can simplify the experimental procedure without the need for any modification to the setup or the control pulse itself.

One strategy to partially compensate for these limitations involves quantum optimal control (QOC)~\cite{Glaser2015, Rembold2020, Mueller2021}. QOC has previously been applied to optimize MW control pulses for quantum sensing with NV centers in a variety of settings~\cite{Rembold2020,Haberle2013,Scheuer_2014,Hernandez2021,Frank2017,Poggiali2018,Mueller2018,Nobauer2015,Ziem2019,Poulsen2021}. Its common objective connects the diverse family of QOC algorithms: to iteratively improve a time-dependent control pulse until a given goal has been reached. Some of these algorithms rely on simulations (open-loop) to quantify the quality of the pulses. In contrast, others achieve the same via direct interaction with the experiment (closed-loop, Fig.~\ref{fig:1}a-c). Algorithms such as GRAPE~\cite{Khaneja2005, Machnes2011} (gradient ascent pulse engineering) or Krotov's method~\cite{Konnov1999,Goerz2019} require the calculation of the derivative of the goal function (gradient-based). The dCRAB algorithm (dressed Chopped RAndom Basis)~\cite{Doria2011,Rach2015,Mueller2021} can be implemented under a gradient-free strategy. Additionally, the functional parametrization approach of the dCRAB algorithm can be combined with gradient search methods via algorithms like GROUP~\cite{Sorensen2019} (gradient optimization using parametrization) or GOAT~\cite{Machnes2018} (gradient optimization of analytic controls). Even with a moderate number of basis functions, the control pulse can contain enough information to steer the system~\cite{Lloyd2014,Mueller2020}.  

\begin{figure}[h]
    \centering
    \includegraphics[width = 0.48\textwidth]{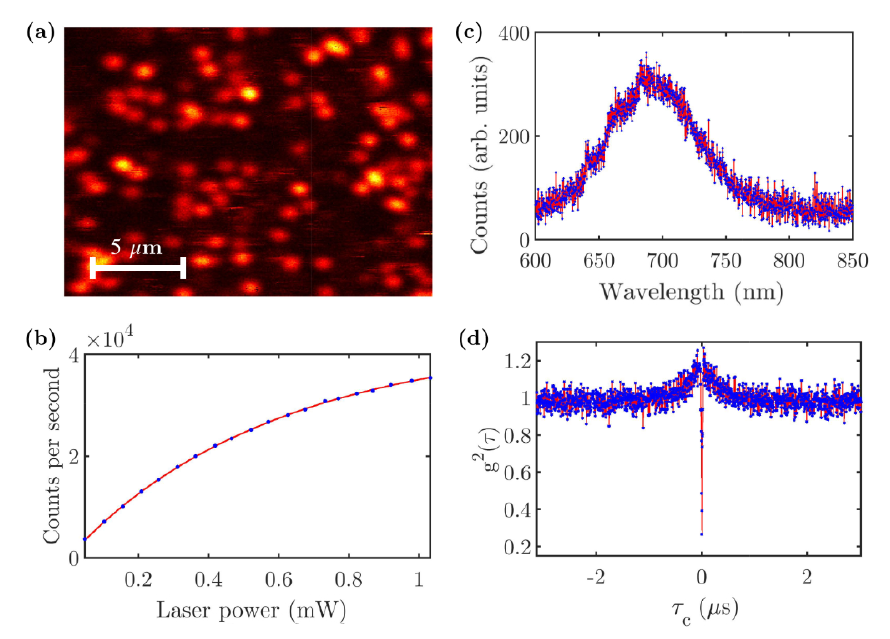}
    \caption{\label{fig:2} Sample characterization; (a) Confocal scan of the diamond sample with shallow single NV centers. (b) Count rate of a single NV vs. the input laser power (this corresponds to a source power in the range of 0 to 40 mW). The saturation behavior can be studied to obtain the excitation power with the best signal-to-background ratio for the experiments. Ideally, this lies below the saturating laser power. (c) Typical emission spectra of the single NV centers in the sample. The NV charge states have different spectral signature, the given spectrum indicates negatively charged NV state. (d) Exemplary second order correlation measurement, which is performed to identify single NV centers in the sample. $\tau_c$ is the delay time in the photon antibunching measurement with the NV ($g_2(0)\approx$ 0.27).}
\end{figure}

With the ultimate goal of enhancing the sensitivity of the main DC magnetometry methods with NV centers (section~\ref{sec:2}), this work presents a two-step strategy to exploit the full potential of feedback-based optimization algorithms and QOC~\cite{Rios2013,Frank2017,Heck2018,Mueller2021} in connection with shallow single NV centers in diamond (<~10\,nm below the surface, Fig.~\ref{fig:2}a). At the first step, the optical spin initialization/readout processes are optimized via a gradient-free Nelder-Mead search~\cite{Nelder1965} in the parameter landscape corresponding to the properties of the experimental system and setup (section~\ref{sec:3.A}). In the second step, we utilize the gradient-free dCRAB algorithm to optimize the MW pulses for spin state manipulation. The optimization routine is implemented via the QOC software package RedCRAB (Remote dCRAB)~\cite{Frank2017,Heck2018}. The optimized MW controls are developed for two DC magnetometry methods (section~\ref{sec:3.B}), namely the pulsed ODMR sequence~\cite{Dreau2011,Acosta2010} and the Ramsey sensing protocol~\cite{Balasubramanian2008, Childress2006}. Two optimization bases, Fourier~\cite{Caneva2011} and Sigmoid~\cite{Rembold2021}, (see appendix~\ref{appb}) are compared to assess their suitability for the involved methods. All optimizations include a Figure of Merit (FoM, see Fig.~\ref{fig:1}a) based on the optical readout contrast. To include robustness against variation in MW drive strength, the FoMs are adapted to scan over control amplitudes ranging from 100\% to 20\% of the maximum. Finally, the optimized pulses are assessed for their enhancement of the average sensitivity and robustness (section~\ref{sec:4}).  
\section{\label{sec:2}DC Magnetometry Methods}
The transitions in the energy level structure of the NV center strongly influence its sensitivity towards external magnetic fields (sensitivity is defined in appendix~\ref{appd}, more details in section~\ref{sec:3.A}). 
The optical ground state forms a spin one triplet system, with a Zero-Field Splitting (ZFS) of $\approx 2.871$\,GHz. In the presence of an external magnetic field along the NV center's axis ($B_\text{NV}$), Zeeman splitting lifts the degeneracy between the $m_s = \pm1$ states. This splitting provides a direct way to quantify $B_\text{NV}$. A pseudo two-level system can be constructed from the $m_s=0$ and one of the $m_s=\pm1$ states. The two-level approximation forms the basis for various magnetometry techniques with NV centers~\cite{Jelezko2006,Childress2006,Taylor2008,Degen2008}.

The most straightforward procedure to detect DC magnetic fields is called continuous wave optically detected magnetic resonance (cw-ODMR)~\cite{Dreau2011, Acosta2009}. The method involves continuous polarization of the NV spin state with a green laser, while MW pulses with different drive frequencies $\omega_\text{mw}$ are applied sequentially to locate the resonance peaks. The splitting between the resonance peaks is proportional to $B_\text{NV}$. Cw-ODMR measurements are less demanding in terms of practical resources and complexity than pulsed measurement schemes, as they do not require pulsed controls. However, by nature, continuous-wave measurements have a lower spin readout fidelity and suffer from optical and MW power broadening~\cite{Dreau2011}.

The dephasing time $T_2^\ast$ sets a limit to the achievable sensitivities with different DC magnetometry methods (see appendix~\ref{appd}, Eq.~\eqref{eq:D4}). Short laser and MW pulses help to overcome the power broadening effect~\cite{Dreau2011} and attain better sensitivities. Pulsed ODMR involves pulsed optical excitations and spin state transfer using MW $\pi$-pulses. For shallow NV centers, the spin states decay quickly. Hence, pulsed ODMR experiments with short, high power control pulses can be advantageous. The short control pulses inherently result in faster measurements, which lead to an improvement in the overall sensitivity. The pulsed ODMR method also offers enhanced readout contrast, which further improves the sensitivity. Note that the sensitivity is defined as the least detectable magnetic field within a measurement time of one second~\cite{Degen2017,budker2020}.

In general, the cw- and pulsed ODMR methods do not exploit the quantum property of spin superposition, which provides a way to make the measurements more sensitive~\cite{Degen2017}. Conversely, the double-pulse-based Ramsey sequence does utilize spin superposition states for sensing. It also has the advantage of avoiding the power broadening effects~\cite{budker_atomic_2008}. The Ramsey method consists of two $\frac{\pi}{2}$-pulses, with free precession time $\tau$ in between. The optically initialized NV spin state is transferred into a superposition state by the first of the two $\frac{\pi}{2}$-pulses. This superposition state interacts with the external magnetic field for the time $\tau$, thus accumulating a phase. Eventually, the second $\frac{\pi}{2}$-pulse converts the accumulated phase into an optically measurable population difference. In contrast to the ODMR-based frequency-sweep methods, the Ramsey sensing protocol is performed at a fixed $\omega_\text{mw}$. In addition, $\tau$ can be varied to measure minimal fluctuations in external magnetic fields~\cite{Degen2017}. In general, the Ramsey method can be used to sense any magnetic fields that change slowly enough, i.e., with frequencies less than $\frac{1}{\tau}$ (see appendix~\ref{appd}).

\section{Sample and Experimental Setup}
All experiments in this work involve an electronic-grade diamond sample (300$\mu$m $\times 100\mu$m $\times$40$\mu$m) with implanted NV centers (Fig.~\ref{fig:2}). The nitrogen ion implantation was performed with a fluence of $3 \times 10^{11}\,$cm$^{-2}$ at 6\,keV, which results in an average depth on around 9.3 $\pm$ 3.6 nm~\cite{bernadi2017}. The implantation was followed by annealing (850$^\circ$C), which forms NV centers in the implanted layer, which shows a uniform density over the sample. Afterward, a second oxidation annealing at 400$^\circ$C was performed, followed by tri-acid cleaning. This process removes the top layer of the diamond, resulting in reduced the NV density and depth. The average NV density is estimated via confocal fluorescence maps to be around 7$\times 10^7 \,$cm$^{-2}$. The value is obtained by analyzing confocal scans of the sample surface (see Fig.~\ref{fig:2}a).  

For the experiment, the sample is mounted on an $\Omega$-shaped strip-line MW antenna (Fig.~\ref{fig:1}c, A)~\cite{Opaluch2021}. The antenna is mounted on a piezoelectric scanner to perform multi-axial scans. The dichroic mirrors (Fig.~\ref{fig:1}c, $D_1$, and $D_2$) filter the excitation laser pulse and direct the fluorescence along the collection arm of the confocal setup. Additionally, a 600\,nm long-pass filter (Fig.~\ref{fig:1}c, F) in the collection arm is used for spectral filtering. The optical initialization and readout are assisted by an objective (Fig.~\ref{fig:1}c, O), which delivers and collects the light to/from the in-focus diamond sample containing shallow single NV centers. Laser pulsing is achieved with a digitally modulated diode laser (modulation bandwidth: 125 MHz). The MW control pulses are generated by mixing the in-phase (I) and quadrature (Q) components (Fig.~\ref{fig:1}b) with a carrier signal. The resulting pulse is subsequently amplified and delivered to the confocal setup via the strip-line antenna. For more details on the setup, see appendix~\ref{appa}.  

\begin{figure*}
    \centering
    \includegraphics[width = 0.99\textwidth]{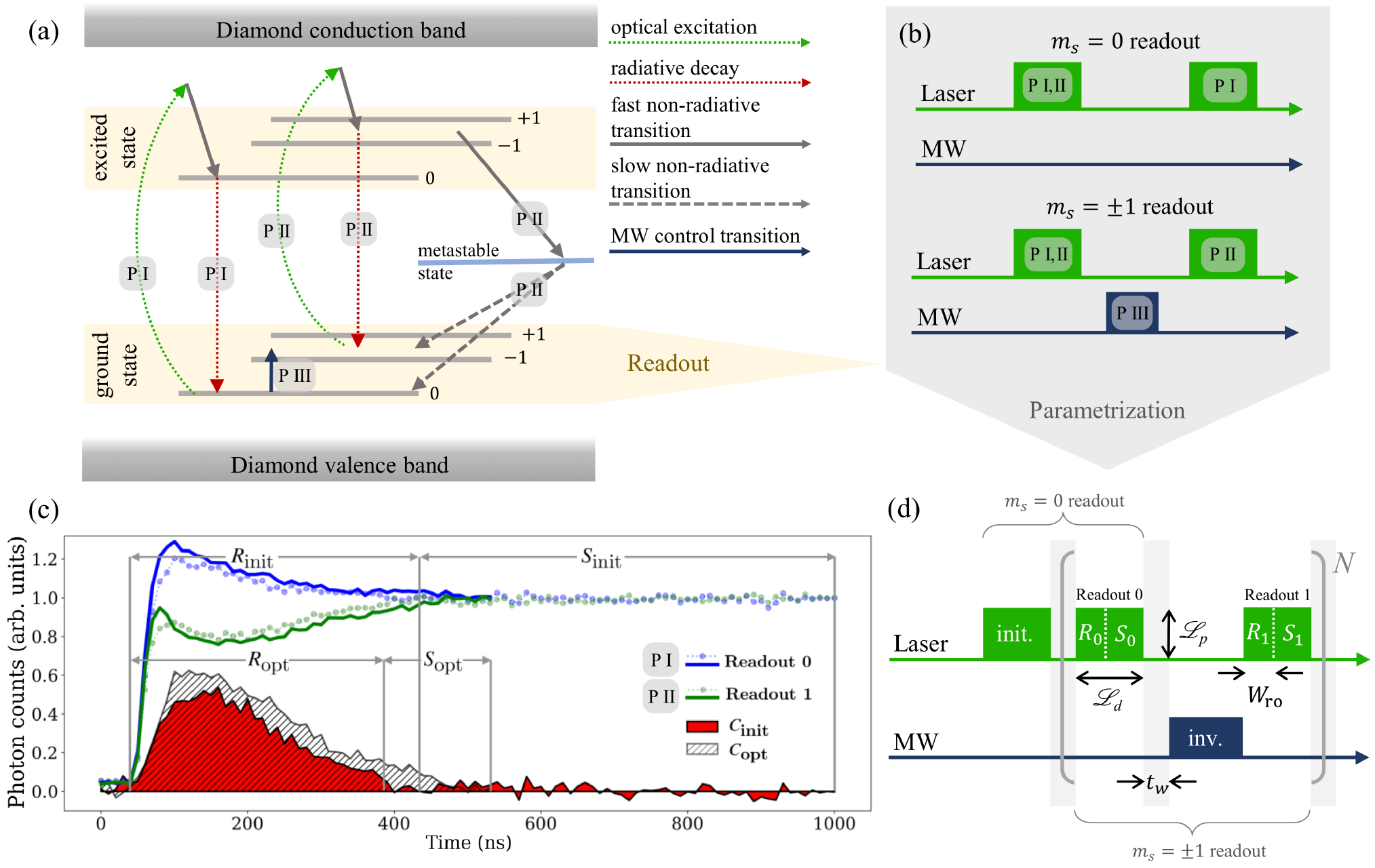}
    \caption{\label{fig:3} Initialization and readout of the NV spin state. (a) The energy level structure of the NV center within the diamond bandgap. The transitions of an NV initially in the $m_s=0$ and $m_s=\pm1$ ground state are denoted as Path I and Path II, respectively. Note that Path II includes a decay via the metastable state, making it slower. A resonant MW pulse, may drive the $m_s=0\leftrightarrow\pm1$ ground state transition (Path III). (b) Spin state readout sequences. By sweeping the MW frequency $\omega_\text{mw}$ this corresponds to a pulsed ODMR sequence. (c) The light curves with dots show a typical readout signal for a 1$\mu$s laser pulse for different initial spin states (blue: $m_s = 0$, green: $m_s = \pm1$). The solid curves indicate the optimized spin state readout (see section~\ref{sec:3.A}). The shaded areas give the readout contrast (Eq.~\eqref{eq:2}) obtained with the 1$\mu$s laser pulse ($C_\text{init}$, red) and the optimized laser pulse ($C_\text{opt}$, striped). $R_\text{opt}$ and $S_\text{opt}$ indicate the optimized windows for the readout and saturation, respectively. (d) Parameters for the optical readout optimization. $R_i$ and $S_i$ correspond to the photon collection windows described in Eq.~\eqref{eq:5}. A spin inverting rectangular MW pulse (inv.) is used for the parameter optimization. Readout 0~(1) corresponds to the readout of the $m_s = 0\,(\pm 1)$ spin states. After an initial laser pulse (init.), each measurement is repeated $N$ times to enhance the signal-to-noise ratio.}
\end{figure*}
In principle, the implementation of these magnetometry methods with NV centers is straightforward and well understood. In practice, however, various factors may affect the performance of these sensing schemes. For example, custom-built MW antennas with unknown instrument response functions are often used to deliver the control pulses. In such cases, the control pulses delivered to the NV center may slightly defer from their actual design. Additionally, for shallow NV centers it is difficult to model all surface effects with adequate precision. This lack of information a priori makes it a challenge to accurately model the system. A closed-loop optimization circumvents this issue.

\section{\label{sec:3}Optimization Methods}

The sensing protocols described in section~\ref{sec:2} rely on the efficiency of two types of control: Readout/initialization via the laser and spin manipulation via the MW field. Here, two complementary optimization strategies are presented using the RedCRAB optimization suite. The first adapts the laser pulse parameters (section~\ref{sec:3.A}) and the second the MW control pulses (section~\ref{sec:3.B}). 
The cloud-based optimization incorporates re-evaluations according to the accuracy of the measurements as well as restrictions of the parameters (see section~\ref{sec:4}), pulses, and superparameters (see section~\ref{sec:3.B}) chosen to compliment the experimental hardware.

In both optimization steps, we first quantify the goal with an FoM that can be measured in the experiment. Subsequently, the controllable constant parameters and time-dependent controls of the system are identified. The initialization/readout is optimized with a direct search, while the MW pulses are optimized via the dCRAB algorithm~\cite{Mueller2021,Doria2011,Rach2015,Heck2018}.
\subsection{\label{sec:3.A}Parameter optimization for Spin State Initialization and Readout}
Strong spin polarization and spin state dependent fluorescence are fundamental to the readout of single NV centers. These properties primarily originate from the transition rates of the spin-preserving radiative and the non-radiative decay channels between the NV energy levels (see Fig.~\ref{fig:3}a). The non-radiative inter-system crossing via the metastable state does not preserve the spin state~\cite{Choi2012}. 
Figure~\ref{fig:3}a shows how an NV, which is originally in $m_s=0$ or $m_s=\pm1$, decays via Path I or II, respectively after being excited by a green laser pulse ($\lambda = 520-530$\,nm). The excited $m_s=0$ state decays radiatively to the ground state, while the $m_s=\pm1$ state might take the non-radiative route via Path II. If the laser pulse is long enough, all population ends up in the $m_S=0$ ground state.\\
Figure~\ref{fig:3}b shows the readout procedure. To obtain a contrast, a laser pulse first initializes the system to the ground state $m_s=0$ via Paths I and II. In the top part of Fig.~\ref{fig:3}b no MW is applied ($m_s=0$ readout) and a second laser pulse leads to a decay via Path I. An intermediate MW pulse ($m_s=\pm1$ readout) can transfer the spin state to $m_s\pm1$ via Path III. The subsequent laser pulse induces a decay via Path II which leads to a drop in fluorescence (Fig.~\ref{fig:3}c) because of the decay via the long-lived metastable state. Hence, the photon count allows differentiating the spin states during optical readout.\\
A simulation-based (open-loop) optimization for optical spin state initialization and readout can be done considering the NV rate equations with experimentally obtained transition rates~\cite{Robledo_2011}. Such methods may require specialized apparatus for optical pulse shaping~\cite{Liu2021}, and in general, do not account for experimental limitations. It is noteworthy that pulse width induced polarization enhancement methods have been previously studied for improving the for spin state initialization using short laser pulses~\cite{Song2020}. In comparison, closed-loop parameter search offers straightforward enhancement. In the following, we develop the method for an optical readout with laser pulses of constant power and finite duration, subjecting to the limitation of the setup.
The photon shot noise is the primary limitation to an efficient optical readout of the NV spin state. Consequently, the statistical determination of the spin state requires an averaged readout over a large number of experimental repetitions. The spin state readout fidelity $\mathcal{F}$ for such probabilistic measurements is expressed in terms of the noise parameter $\sigma_R$~\cite{Taylor2008, Shields2015}:
\begin{equation}
\label{eq:1}
    \frac{1}{\mathcal{F}}=\sigma _{R}\approx\sqrt{1+\frac{2\left( R_{0}+R_{1}\right)}{\left( R_{0}-R_{1}\right) ^{2}}},
\end{equation}
such that $\mathcal{F} = 1$ at the spin projection noise limit of the sensitivity (see appendix~\ref{appd}). $R_1$ ($R_0$) is the total number of collected photons from the readout of the spin state initialized in $m_s=\pm1$ ($m_s=0$). Experimentally, the readout contrast $C$ is given by
\begin{equation}
\label{eq:2}
    C = \frac{R_0 - R_1}{R_0 + R_1}.
\end{equation}
Its relation to $\mathcal{F}$ is given in appendix~\ref{appd}.
Intrinsically, the contrast depends on several system properties and experimental parameters,
\begin{equation}
    \label{eq:3}
    C \equiv C[\gamma_{ij},\mathscr{L}_p,\mathscr{L}_d,\Omega_{\text{max}}, B_{\perp},E_{xy}, T....],
\end{equation}%
where $\gamma_{ij}$ is the transition rate between levels $i\leftrightarrow j$, $\mathscr{L}_p$ is the laser pulse intensity, $\mathscr{L}_d$ is the laser pulse duration, $\Omega_{\text{max}}$ corresponds to the maximum amplitude of the spin inversion control pulse, $B_\perp$ and $E_{xy}$ are off-axial magnetic and electric field components at the position of the NV center respectively, and $T$ is the ambient temperature. In addition, several other factors, including crystal field strain and charge state stability, may affect the fluorescence of the NV center and ultimately influence the readout contrast. The majority of the parameters in Eq.~\eqref{eq:3} depend on the system properties, material characteristics, and ambient conditions that are generally not fully controllable. In practice, some of the system properties can be characterized before the optimization of the readout contrast. For example, the charge state of the NV center can be determined from the emission spectrum (Fig.~\ref{fig:2}b). Similarly, external factors such as crystal field strain and temperature directly influence the ZFS of the NV center. In this regard, pre-characterized single NV centers (Fig.~\ref{fig:2}c) with ZFS~$\approx 2.871$\,GHz, and stable photoluminescence that do not exhibit charge state related blinking allow to fully exploit the scope of laser pulse parameter optimization (Note that this does not rule out photochromism on short timescale). Likewise, a well-aligned static magnetic field $B_\text{NV}$ is a prerequisite for the optimizations performed in presence of a magnetic bias field. It is noteworthy that photons originating from NV$^0$ can be filtered from the readout signal (Fig.~\ref{fig:1}a). As a result, charge state instability leads to blinking of the NV fluorescence signal~\cite{Aslam2013}. 

Other experimental parameters in Eq.~\eqref{eq:3} such as $\mathscr{L}_p$ and $\mathscr{L}_d$, directly influence the optically induced transitions, as well as the charge state stability~\cite{Shields2015, Aslam2013}. In contrast, the effect of the wait time $t_w$ between the initialization and readout/spin manipulation pulses (shaded region between pulses in (see Fig.~\ref{fig:3}d)) is more indirect. Hence, it is commonly set to ca. $300\,$ns, which corresponds to the lifetime of the metastable state~\cite{rendler2018}. Similarly, the photon collection window $W_\text{ro}$ is often calculated in advance to obtain the best SNR for every readout~\cite{Doherty2013}.
Consequently, Eq.~\eqref{eq:3} can be reduced to a simpler form based on the variables that can be controlled experimentally,
\begin{equation}
\label{eq:4}
    C \thicksim C[\mathscr{L}_p,\mathscr{L}_d,W_\text{ro},t_w,\Omega_{\text{max}}].
\end{equation}

Although it is not straightforward to find an analytical form to characterize the dependence of $C$ on these parameters, they can be directly adjusted in a closed-loop optimization on the experiment. Figure~\ref{fig:3} shows the two-shot scheme for the contrast measurement used in the optimization routine. This strategy is devised keeping in mind that the experimental setup does not enable time-tagged photon counting. Each laser pulse (Readout 0 (1)) is divided into a spin readout window $R_0$ ($R_1$), and a spin state saturation window $S_0$ ($S_1$). Their durations are determined by the optimization parameters $\mathscr{L}_d$ and $W_\text{ro}$. A spin inversion MW pulse flips the spin state between the laser pulses. The FoM, which is minimized during the optimization, is given by
\begin{equation}
\label{eq:5}
        \text{FoM}_{\text{RO}}=1 -\bar{C}\left[ 1-\text{var}\left( \begin{vmatrix}
    S_{0}-S_{1} \\
    \overline{S_{0}+S_{1}}
    \end{vmatrix}\right) \right].
\end{equation}
Here, $\bar{C}$ is the readout contrast averaged over $N$ experimental repetitions as shown in Fig.~\ref{fig:3}d. In addition to maximizing the readout contrast in the spin readout windows, $\text{FoM}_{\text{RO}}$ also ensures uniform spin state initialization, as the optimized value tends to minimize the variance in the photon counts from the two spin states in the saturation windows (Fig.~\ref{fig:3}).
The closed-loop optimization of readout parameters is generally relevant for a variety of methods, such as readout based on spin-to-charge-state conversion~\cite{Shields2015} and photoelectric readout~\cite{bourgeois2015}, which inherently involves laser pulses. Furthermore, the optimized readout can be integrated directly with MW-free, all-optical magnetometry methods~\cite{Tetienne_2012}. The standard approach of obtaining the optimal readout parameters usually requires multiple measurements, or otherwise time-tagged photon counting. The closed-loop optimization approach presented reduces the measurement time that otherwise will be required to manually optimize all the parameters under consideration. This is further improved by reducing the measurement to two readouts only for assessing $R_i$ and $S_i$.
\subsection{\label{sec:3.B}Quantum Optimal Control for Spin State Manipulation}
An optimally initialized spin state and its efficient readout are two of the essential criteria for a practical quantum sensor~\cite{Degen2017}. In addition, the spin state has to be controlled accurately to implement a sensing protocol. Following parts of the text describe the optimization of MW control pulses for spin inversion and for a $\left(\frac{\pi}{2}\right)_x$-gate via the dCRAB algorithm~\cite{Mueller2021,Doria2011,Rach2015,Heck2018}.

Before proceeding to the specifics of the optimization schemes, we discuss the dynamical equations of the system to introduce the basic concept of QOC. The system is described by a constant drift Hamiltonian $H_d$, and control Hamiltonians $H_c^i$, which are modulated by control pulses $u^i(t)$:
\begin{equation}
    \begin{split}
        H(t)&=H_d + \sum_i H_c^i u^i(t)\\
        &=\frac{\hbar}{2}\left(\Delta \sigma_z + \sigma_x u^1(t) + \sigma_y u^2(t)\right),
    \end{split}
\end{equation}
where the complete Hamiltonian $H(t)$ is given in the rotating wave approximation (RWA) with the detuning $\Delta = \omega_\text{mw}-\omega_\text{nv}$, the NV's resonant frequency $\omega_\text{nv}$, the Pauli matrices $\sigma_i$, and the controls $u^1(t)=\Omega(t) \cos(\phi(t))$ and  $u^2(t)=\Omega(t) \sin(\phi(t))$. These controls correspond to the in-phase and quadrature components of a MW drive, with Rabi frequency $\Omega(t)\in [0,\Omega_\text{max}]$ and phase $\phi(t)$ applied for the duration $t_p$.

The control objective for the MW pulses is to efficiently transfer the initial spin state $\ket{\Psi_i}$ to the final state $\ket{\Psi_f}$. Hence, the FoM is defined as the state fidelity,
\begin{equation}
\label{eq:7}
    \mathcal{F}_p = \lvert\bra{\Psi_f}U(t)\ket{\Psi_i}\rvert^2,
\end{equation}
\begin{equation}
\label{eq:8}
    \text{where }U(t) = \mathcal{T}\text{exp}\left[{-\frac{i}{\hbar}\int_{0}^{t_p}H(t)dt}\right],
\end{equation}
where $\mathcal{T}$ indicates a time-ordered exponential propagator. At this point, the FoM is a functional of the control pulses. The controls are subsequently parametrized by a set of $N_\text{set}\times M$ basis elements $f^i(\omega_n;t)$. Each element is defined by its superparameter $\omega_n$, which is randomly selected from $\omega_\text{min}<\omega_n<\omega_\text{max}$, where $\omega_\text{min}$ and $\omega_\text{max}$ are the minimum and maximum allowed values. The number of basis functions $M$ per superparameter depends on the basis. These superparameters can be the frequencies of a set of trigonometric functions (Fourier basis~\cite{Rach2015}; in this case $\omega_\text{min}$ and $\omega_\text{max}$ set the allowed bandwidth of the control pulse) or the offsets for a set of step functions (Sigmoid basis~\cite{Rembold2021}). The resulting pulses take the following form:
\begin{equation}
    u^i(t)=u_0^i(t) + \sum_n^{N_\text{set}} \sum_i^M A_n f^i(\omega_n;t).
\end{equation}
Here, $u_0^i(t)$ represents the initial guess for the pulse.

Following the parametrization, the goal of the QOC routine is to find the optimal values for the coefficients $A_n$, maximizing the FoM (Eq.~\eqref{eq:5}). Especially in closed-loop optimization, only a limited number of parameters can be optimized at a given time. Therefore, additional steps are required to avoid local optima. The dCRAB algorithm tackles this issue by switching the set of basis elements every time the optimization has converged under the given constraints~\cite{Rach2015}. Every new optimization (superiteration) is started with the previous optimum as an initial guess, i.e. $u_0^i(t)=u_\text{opti}^i(t)$.

\begin{figure}[b]
    \centering
    \includegraphics[width=0.4\textwidth]{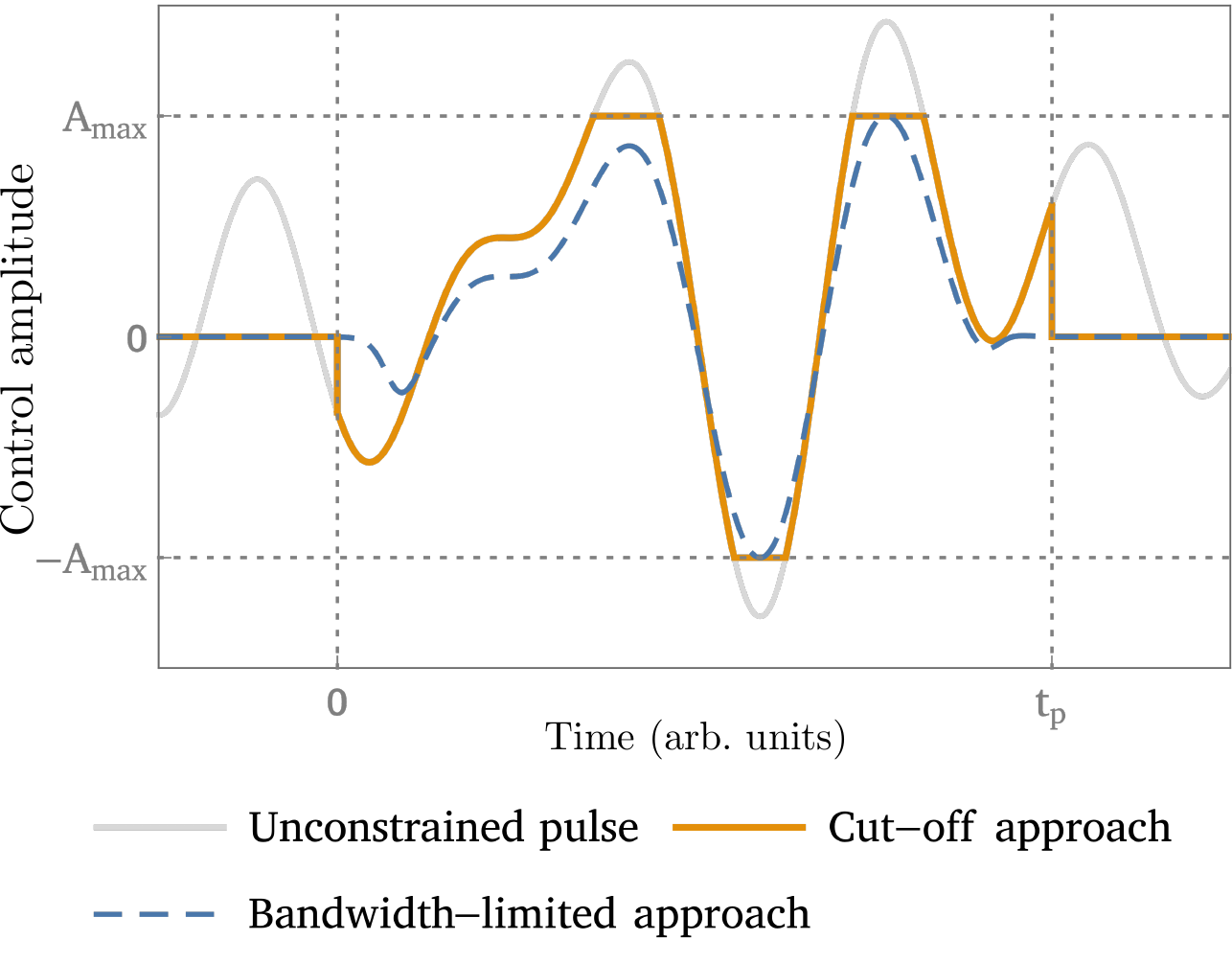}
    \caption{ Restriction approaches. The unconstrained pulse may either be cut off at the amplitude $\{-A_\text{max},A_\text{max}\}$ and time $\{0,t_\text{p}\}$ limits ("cut-off approach") or shifted and rescaled to fit within the available window ("bandwidth-limited approach").}
    \label{fig:5}
\end{figure}
The optimizations are performed with both, the Fourier and Sigmoid basis separately. To ensure the pulse amplitude and duration are limited, i.e., that the amplitude stays within an upper and a lower limit and the pulse is zero at $t=0$ and $t=t_p$, two different strategies are applied and illustrated in Fig.~\ref{fig:5}. In the cut-off approach, the pulses are cut off at $t=0$ and the $t=t_p$ to limit the duration. Similarly, they are cut off at the top and bottom to force the amplitude limits. Instead, the bandwidth-limited approach involves re-scaling the pulse to fit within the amplitude limits, followed by multiplication with a smooth window function like a flat-top Gaussian to avoid discontinuities at initial and final time.

In the cut-off approach (see Fig.~\ref{fig:5}), the Fourier basis is expected to produce high-frequency components when the optimization algorithm maximizes the pulse area. Conversely, in the bandwidth-limited approach, the Fourier basis will have difficulties to significantly expand the pulse area. At the same time, the Sigmoid basis has the ability to exploit the pulse area without producing high frequencies~\cite{Rembold2021} when combined with the bandwidth-limited approach. The inherent smoothness offered by the Sigmoid basis (see appendix~\ref{appb}) provides a particular advantage for frequency-sweep-based spectroscopic measurements, where spurious harmonics are to be avoided. 

The first MW optimization presented here concerns the spin-inversion pulse in the pulsed ODMR sequence (see Fig.~\ref{fig:3}). The efficiency of the spin state transfer is estimated through the optical readout contrast $C$ (Eq.~\eqref{eq:3}). The previously obtained parameters for the laser-based initialization/readout are used as the default for the MW control pulse optimization experiments. To achieve robustness, the control field amplitude variation is incorporated in the FoM by averaging the contrast over a range of Rabi frequencies $\Omega_\text{max}$.
\begin{equation}
\label{eq:10}
    \text{FoM}_{\text{podmr}}=1-\frac{1}{N_p}\sum_k^{N_p}\left( \frac{R_{0}^{k}-R_{1}^{k}}{R_{0}^{k}+R_{1}^{k}}\right),
\end{equation}
where $N_p$ is the total number of sampled $\Omega_\text{max}$ and $R_{i}^{k}$ are the photon counts from the corresponding spin state collected during Readout 0 (1) (see Fig.~\ref{fig:3}). The goal of the optimization is to minimize $\text{FoM}_{\text{podmr}}$.

\begin{figure}
    \centering
    \includegraphics[width = 0.45\textwidth]{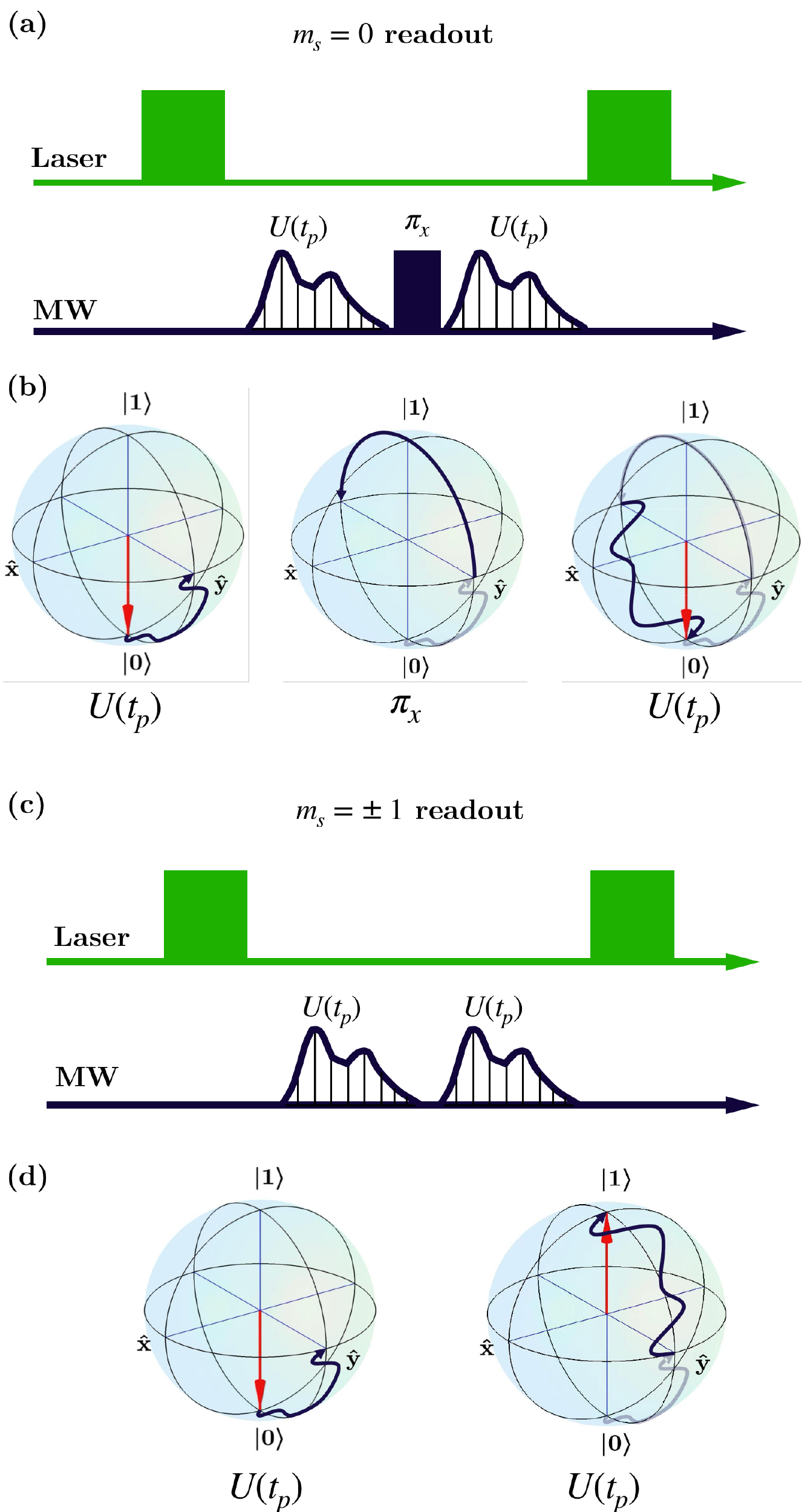}
    \caption{\label{fig:4} Exemplary measurement protocol for the Ramsey sequence optimization. The spin is projected into $m_s=0$ and $m_s=\pm 1$, similarly to the spin state measurement in Fig.~\ref{fig:3}b. (a) $m_s = 0$ state ($\ket{0}$) projection: A known refocusing $\pi_x$-gate (solid) is applied between two optimized pulses $U(t_p)$ (shaded). (b) Exemplary Bloch sphere representation of the process in (a). Red arrows indicate the initial and final spin state, and dark blue lines denote the path of the spin state. (c) In the absence of the intermediate $\pi_x$-pulse, the spin state is ideally transferred to the $m_s = \pm1$ state ($\ket{1}$). (d) Exemplary Bloch sphere representation of the scheme in (c).}
\end{figure}

The Ramsey protocol does not involve spin inversion, but instead a $\left(\frac{\pi}{2}\right)_x$-gate. In the sensing procedure, this pulse plays two roles: First, it maps the spin eigenstates to a superposition state with a given phase. Second, it converts the phase back to a spin population. Gates cannot be directly quantified using the contrast. Instead, their quality is commonly quantified via gate tomography, which requires additional state preparations and related measurements. We develop a protocol to translate the $\left(\frac{\pi}{2}\right)_x$-gate's unitary properties into a readout contrast that takes the same number of measurements as the evaluation of the spin state inversion.
Figure~\ref{fig:4} shows the scheme connecting the pulse performance to the readout fluorescence contrast $C$ from two spin states. Similar to the case of pulsed ODMR, the FoM is defined as  
\begin{equation}
\label{eq:11}
    \text{FoM}_{\text{ram}}=1-\frac{1}{N_p}\sum_k^{N_p}\left( \frac{\mathcal{P}_{0}^{k}-\mathcal{P}_{1}^{k}}{\mathcal{P}_{0}^{k}+\mathcal{P}_{1}^{k}}\right),
\end{equation}
where $\mathcal{P}_i^k$ is the photon count for the $k$th amplitude value after projection into spin state $i$. The photon counts $\mathcal{P}_i^k$ are related to the spin transfer to the different states using the following series of transformations,
\begin{align*}
   U(t_p)\; \pi_x \;U(t_p)&\longmapsto\mathcal{P}_0, \\
   U(t_p) \; U(t_p)&\longmapsto\mathcal{P}_1,
\end{align*}

where $U(t_p)$ is the parametrized unitary operator for the optimized control pulse of duration $t_p$, and $\pi_x$ denotes the unitary transformation for the rectangular $\pi$-pulse applied along the $x$-axis. The maximization of the contrast ideally corresponds to the following conditions:
\begin{align}
    |\langle 0|U(t_p)\; \pi_x \;U(t_p)|0\rangle|^2 = 1\label{eq:pihalf-reversed},\\
    |\langle 1|U(t_p) \; U(t_p)|0\rangle|^2 = 1\label{eq:pihalf-twice}.
\end{align}
Here, $\ket{0}$ and $\ket{1}$ denote the two spin states of the system under the two-level approximation: $\ket{0}$ is given by the $m_s=0$ state and $|1\rangle$ represents either $m_s= +1$ or $m_s=-1$ depending on the corresponding experiment specified in section~\ref{sec:4}. We introduce the parametrization of the unitary transformation generated by the control pulse as
\begin{align}
    U(t_p)=\mathrm{exp}\left[-i\sum_j c_j \sigma_j\right],
\end{align}
with coefficients $c_j$ for $j=\{x,y,z\}$, and $\hat{c}_j=c_j/c$, with $c=\sqrt{c_x^2+c_y^2+c_z^2}$. Then, Eq.~\eqref{eq:pihalf-twice} implies
\begin{align}
    1=\sin^2(2c)(\hat{c}_x^2+\hat{c}_y^2)
\end{align}
and thus $c=\frac{1}{2}\left(\frac{\pi}{2}+k\pi\right)$, for integer $k$, and $c_z=0$. Substituting this into Eq.~\eqref{eq:pihalf-reversed} gives
\begin{align}
    1=4 \hat{c}_x^2 \sin^2c \left[\cos^2c +\hat{c}_z^2\sin^2c\right]=\hat{c}_x^2,
\end{align}
finally indicating that $c_x=c=\frac{1}{2}\left(\frac{\pi}{2}+k\pi\right)$, and hence $c_y=0$. In other words, $\text{FoM}_\text{ram}$ in Eq.~\eqref{eq:11} is minimized for a $\frac{\pi}{2}$ rotation around the $x$-axis (in the positive or negative direction):
\begin{equation}
    U_\text{opti}(t_p)=\exp\left[-\frac{i}{2}\left(\frac{\pi}{2}+k \pi\right)\sigma_x\right].
\end{equation}

\section{\label{sec:4}Experimental results and sensitivity analysis}
A straightforward way to test the general applicability of the optimization strategies discussed in the preceding section is to apply them to different single NV centers and compare the readout contrast enhancement on a case-specific basis. In addition, the average sensitivities from the experiment quantify the optimization benefits. The optimization schemes from section~\ref{sec:3} are implemented and compared in the following section. First, we assess the improvements resulting from optimized readout (OR) (section~\ref{sec:4.A}) and the additionally optimized spin transfer pulses (section~\ref{sec:4.B}) for the pulsed ODMR method. Second, OR is applied with optimized control pulses for the Ramsey protocol, and the results are discussed in section~\ref{sec:4.C}. Finally, the robustness of the pulses is tested over a range varying from 100\% to 10\% of the maximum control power. This variation is artificially introduced in the experiment by changing the power at the MW source.

\subsection{\label{sec:4.A}Initialization and Readout}
Experimental restrictions are directly included in the closed-loop optimization of the initialization and readout process by limiting the optimization parameters. The bounds on the parameter set $\{\mathscr{L}_p,\mathscr{L}_d,W_\text{ro},t_w\}$ are given as:
\begin{align*}
   &\mathscr{L}_p \in \left[2,40\right]\text{   (mW), } \\
   &\mathscr{L}_d \in \left[300,2000\right] \text{   (ns),} \\
   0.25\,\mathscr{L}_d \le\, &W_\text{ro} \le 0.75\,\mathscr{L}_d \text{   (ns)},\\
   \text{and  }&t_w \in \left[0,1000\right]\text{   (ns). }
\end{align*}
\begin{figure}
    \centering
    \includegraphics[width = 0.48\textwidth]{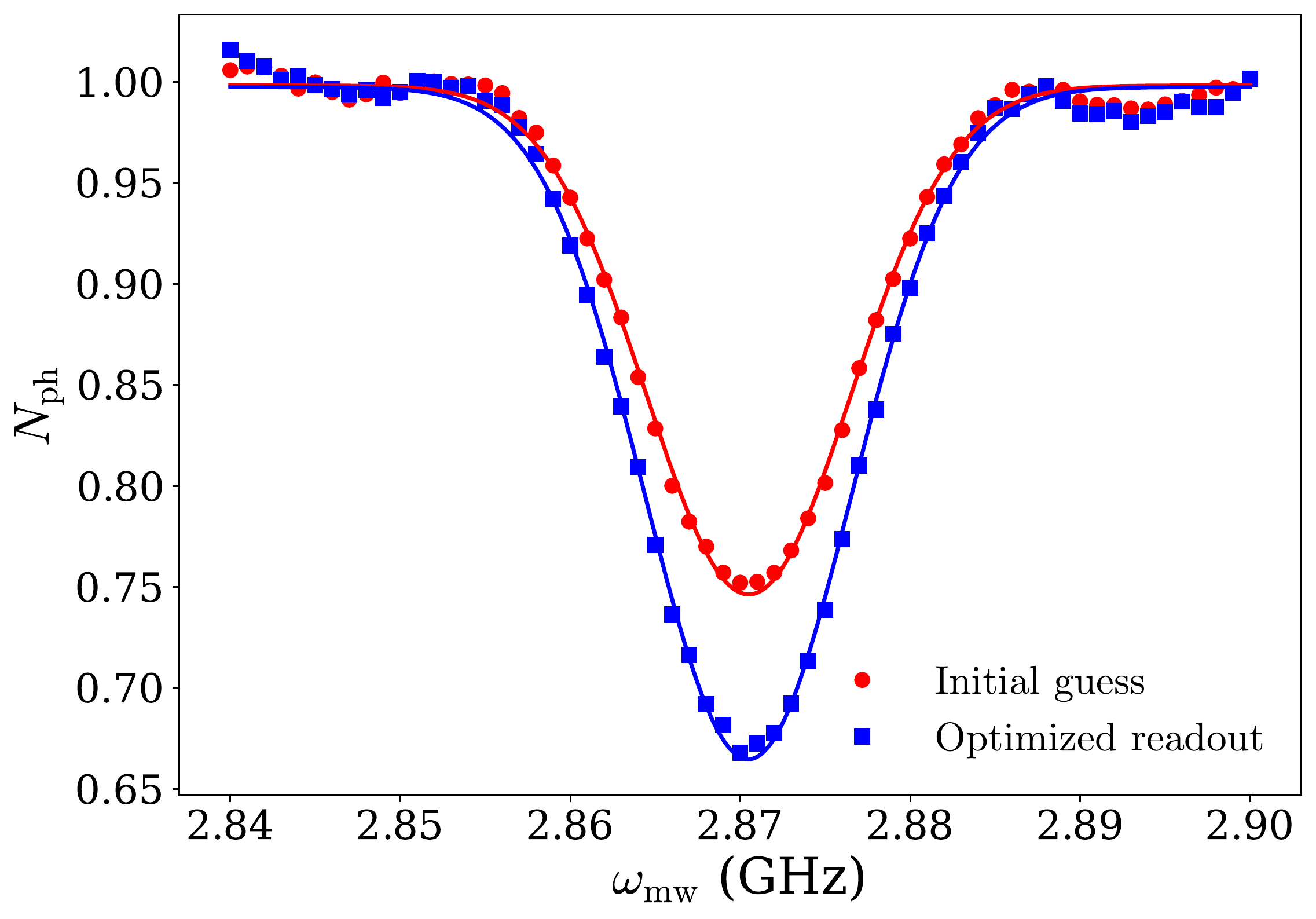}
    \caption{\label{fig:6} Pulsed ODMR at ZFS with optimized parameters using a rectangular MW $\pi$-pulse of the duration of 57\,ns. The experiment with optimized laser parameters exhibits an improved readout contrast $\bar{C}$ of ca. 0.33 (blue) in comparison to the initial guess with a contrast of ca. 0.25 (red).}
\end{figure}
Limits on $\mathscr{L}_p$ correspond to the available source laser power. The incident laser power experienced by the NV center lies between 0.05\,mW and 1\,mW (Fig.~{\ref{fig:2}}(b)). However, as the instrument values are used as a parameter in the optimization, they are the ones referenced throughout the text. The initial guess for the optimization is chosen to be \{$\mathscr{L}_p \le P_{\text{sat}}$, 1000\,ns, 450\,ns, 300\,ns\}, where $P_{\text{sat}}$ is the saturation laser power for the single emitter). In cases where the saturation limit cannot be reached with the available laser intensity, the initial guess is obtained by considering the saturation curve to identify the approximate laser intensity with the most favorable signal-to-background ratio. Some of the readout optimization results are summarized in table~\ref{tab:table1}. As a general observation, the optimized laser pulses are shorter than the corresponding initial guesses, while the $t_w$ values remain almost unchanged after the optimization. Moreover, reduction of the measurement time improves the overall sensitivity of the NV center (Eq.~\eqref{eq:D5}). Figure~\ref{fig:3}c shows the photoluminescence behavior of one of the NV centers involved in the experiment (table~\ref{tab:table1}, NV3). The collected signal reflects the improvement in the average readout contrast after the optimization.

\begin{table}
    \caption{\label{tab:table1} Optimized parameters for spin state readout contrast with single NV centers. Experiments with NV1 are performed at the ZFS, whereas NV2 and NV3 related experiments are performed with a bias field of 12 mT.}
    \begin{ruledtabular}
        \begin{tabular}{lccccc}
            Identifier & $\mathscr{L}_p^{opt}$ & $\mathscr{L}_d^{opt}$& $W_\text{ro}^{opt}$ & $t_w^{opt}$ & Ref.\\
            &[\text{mW}]&[\text{ns}]&[\text{ns}]&[\text{ns}]&\\
            \colrule
            NV1&21&585&260&470& Fig. \ref{fig:6}\\
            NV2&17&488&250&270& \footnote{Pulse optimization restriction via the cut-off approach.}Fig. \ref{fig:7}\\
            NV3&16&552&385&260& \footnote{Pulse optimization restriction via the  bandwidth-limited approach.}Fig. \ref{fig:8}, \\
            &&&&&\ref{fig:9}, \ref{fig:10}, \ref{fig:11}\\
        \end{tabular}
    \end{ruledtabular}
\end{table}

The optimized laser parameters are tested by combining them with a standard pulsed ODMR sequence with rectangular spin inversion pulses (pulse duration of 57\,ns). Their readout contrast is quantified as $\bar{C}=1-\text{min}[N_\text{ph}]$, where $N_\text{ph}$ is the normalized photon count (see appendix~\ref{appd} for details). Figure~\ref{fig:6} shows a comparison between the measurement with and without optimized parameters (zero bias field, table~\ref{tab:table1}, NV2). The optimized parameters account for a 33\% improvement in peak contrast. This result can be improved even further by also optimizing the spin inversion pulses. 
\subsection{\label{sec:4.B}Pulsed ODMR measurements with optimized MW pulses}
The spin inversion pulse that is part of the pulsed ODMR protocol provides a target for further optimization on top of the optimized optical readout. In this regard, we investigate the additional improvement by optimizing the pulses under a bias field $B_\text{NV}$ to emulate a spin resonance sensing scenario. The FoM is calculated by averaging the contrast over a set of $N_p = 5$ measurements (see Eq.~\eqref{eq:10}) leading to control pulses in the range of 4\%\,-\,100\% of the maximum control power (or equivalently, 20\%\,-\,100\% of the maximum control amplitude, $\Omega_\text{max}$). The initial guess resembles a standard rectangular $\pi$-pulse, whose length is determined by observing Rabi oscillations.

\begin{figure}
    \centering
    \includegraphics[width = 0.48\textwidth]{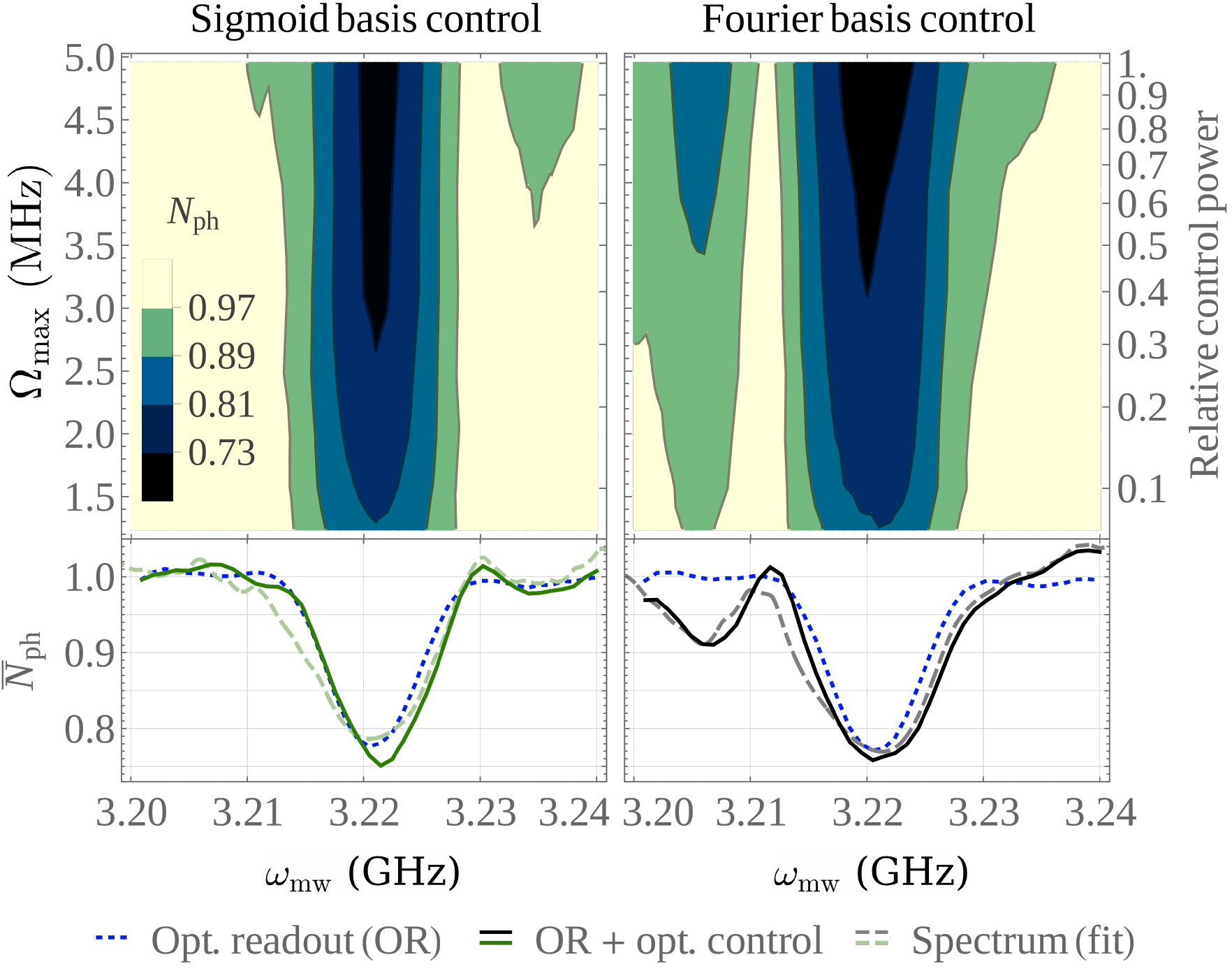}
    \caption{\label{fig:7} Comparison between two optimized spin inversion pulses in presence of a bias field $B_\text{NV}=12\,$mT (table~\ref{tab:table1}, NV2). The left (right) side shows the experimental results from a pulse optimized with the Sigmoid (Fourier) basis. (\textit{top})~Normalized counts $N_\text{ph}$ over a range of $\Omega_\text{max}$ and drive frequencies $\omega_\text{mw}$. (\textit{bottom})~Average normalized count $\bar{N}_\text{ph}$ over all $\Omega_\text{max}$ for the optimized pulse (solid) and initial guess (blue, dotted). The spectra of the pulses (dashed) are convoluted with the NV's natural emission line and fitted to the average counts.}
\end{figure}

Figure~\ref{fig:7} shows two maps representing the normalized count obtained with two optimized MW pulses. In this example, the laser pulses were pre-optimized according to the method described in~\ref{sec:4.A}, and the $m_s = 0\leftrightarrow+1$ transition is used for the optimization as well as the assessment via pulsed ODMR. The MW pulses are optimized according to Eq.~\eqref{eq:10} at the center frequency of 3.22\,GHz. The pulse corresponding to the left is optimized with the Fourier basis, while the right pulse is optimized with the Sigmoid basis, both with a pulse duration of 200\,ns. Both pulses exhibit robustness with respect to the amplitude variations, improving the contrast compared to the initial guess. However, the Sigmoid basis pulse is spectrally narrow, while the Fourier pulse has a distinct sideband. The spectral shape of the pulses can explain these features. The Fourier basis contains high-frequency elements caused by the cut-off limitation (see Fig.~\ref{fig:5}), which the Sigmoid basis avoided. The small off-resonant area addressed by the Sigmoid basis covers only a fraction of the Fourier basis' sideband and is significantly weaker. This is illustrated in the average plot at the bottom.

\begin{figure}
    \centering
    \includegraphics[width = 0.48\textwidth]{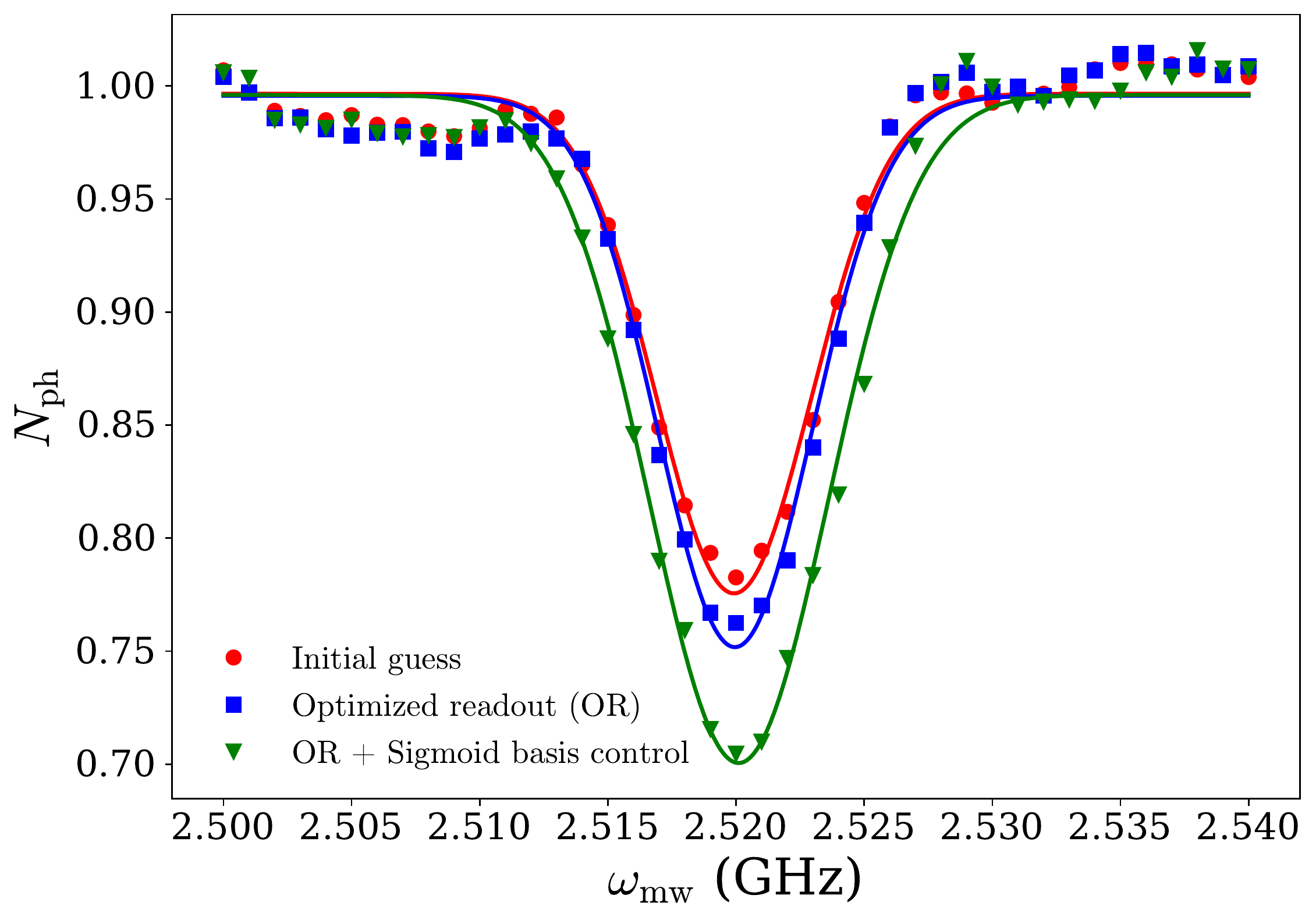}
    \caption{\label{fig:8} Pulsed ODMR in presence of a bias magnetic field ($B_\text{NV} \approx$ 12\,mT, NV3 from table~\ref{tab:table1}) with optimized laser parameters and MW pulses. Following the results from Fig.~\ref{fig:7} the optimization was done with the Sigmoid basis, using the bandwidth-limited restriction approach. The data shows the improvement in contrast with each step of the optimization. The initial contrast for the resonance peaks is ca. 0.22 (initial guess, red), which is further improved to ca. 0.24 with optimized laser parameters (blue). The MW pulse optimized in the Sigmoid basis on top improves the contrast to ca. 0.30 (green). the solid lines show the Gaussian fits for the respective data (see appendix~\ref{appd}).}
\end{figure}

To test the general applicability of this method for generating bandwidth-limited control pulses, similar optimization and pulsed ODMR experiments are performed with a different NV center (table~\ref{tab:table1}, NV3), this time using the $m_s =0\leftrightarrow \text{-}1$ transition, and the bandwidth-limited approach (Fig.~\ref{fig:5}). The results are shown in Fig.~\ref{fig:8}. Here, a readout contrast of ca. 0.24 is obtained with optimized laser parameters (pulse duration of 134\,ns, obtained from observing Rabi oscillations). The Sigmoid pulse (pulse duration of 200\,ns) enhances the readout contrast further to ca. $0.30$. Pulsed ODMR experiments with different peak control power are performed to test the robustness of the control pulse. The readout contrast and Full Width Half Maximum (FWHM) of the resonance profile are obtained by fitting the data with a Gaussian profile (see Eq.~\eqref{eq:c6}).

\begin{figure}
    \centering
    \includegraphics[width = 0.48\textwidth]{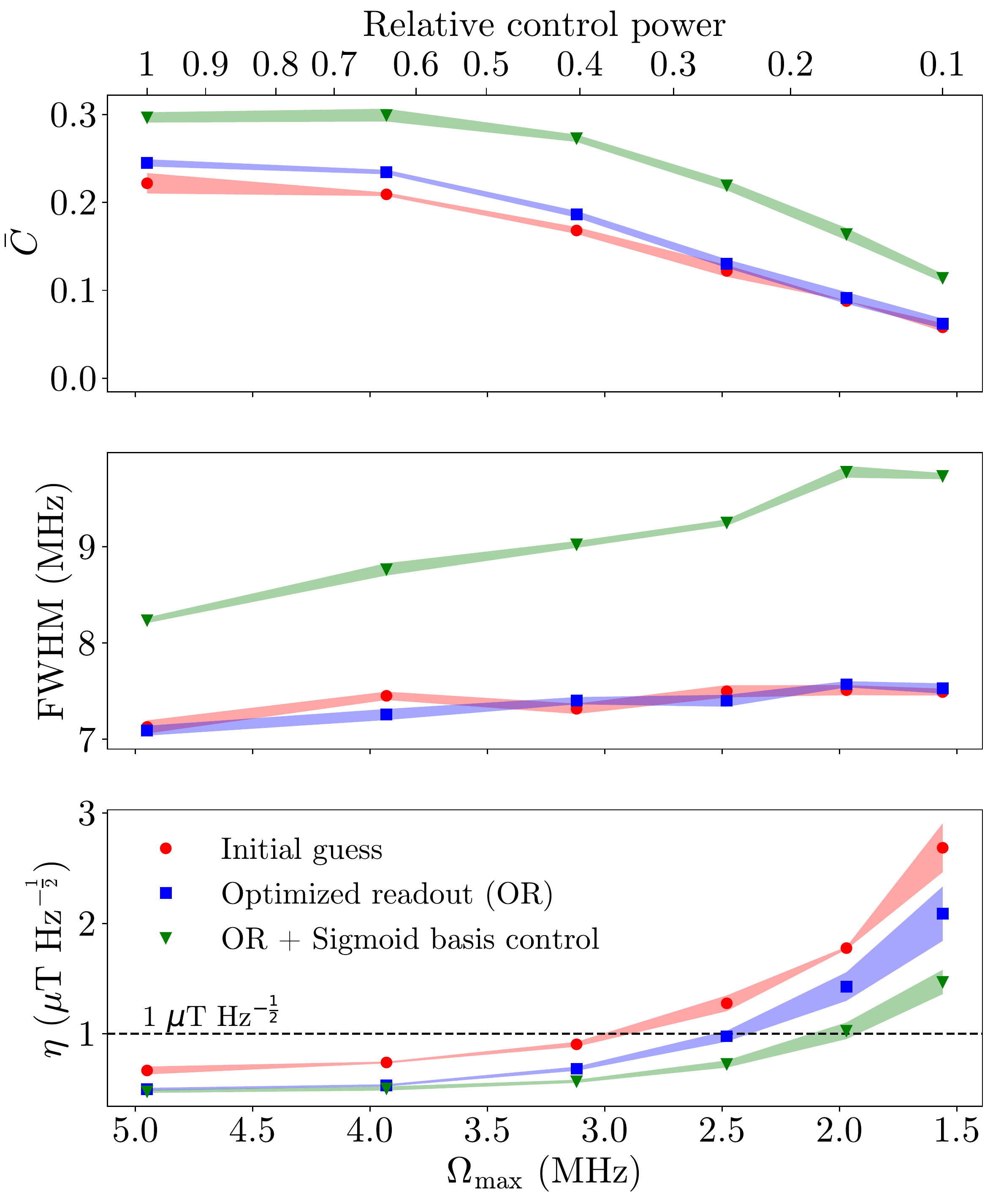}
    \caption{\label{fig:9} Comparison of pulsed ODMR measurements with optimized and standard spin state inversion pulses (NV3 from table~\ref{tab:table1}). Specifically, the robustness against amplitude variation is shown for pulses optimized in the Sigmoid basis. From top to bottom, the contrast, and FWHM of the resonance peaks as well as the corresponding average sensitivity $\eta$ are shown. All results are obtained with different amplitude variations. The red curves indicate the initial guess. Blue curves correspond to the experiments performed with optimized spin readout parameters. The green curves show the results for the experiments using optimized MW pulses. The dashed line in the bottom plot shows the ceiling for the $\eta$ of 1$\mu$T\,Hz$^{-\frac{1}{2}}$. }
\end{figure}

Figure~\ref{fig:9} shows the achievable average sensitivity $\eta$ of the pulsed ODMR method. It depends on the resonance profile, its FWHM, contrast, and the measurement time involved in the experiment (see Eq.~\eqref{eq:D5}). In addition, the spin-projection noise sets a lower limit to $\eta$. The full optimization, including the laser parameters and the robust Sigmoid pulse, leads to a sub-$\mu$T\,Hz$^{-\frac{1}{2}}$ average sensitivity considering up to almost 83\% variation in the control power (see appendix~\ref{appd} for details on the sensitivity calculation). The lack of straight-forward interpretability of solutions lies in the nature of QOC. Still, it is notable that the Sigmoid pulse's FWHM is broader for lower $\Omega$, possibly hinting at the algorithm compensating for the reduced coverage of the hyperfine lines in that regime.

Off-axial magnetic field components lead to spin-mixing, reducing the readout contrast~\cite{Tetienne_2012}. This effect becomes apparent when comparing the contrast at ZFS (Fig.~\ref{fig:6}) and in presence of an external magnetic field (Fig.~\ref{fig:8}). 
The degree of spin-mixing and its effects on the transition rates cannot be straightforwardly simulated for the presented experiments. Using closed-loop optimization of the laser pulse parameters allows to nevertheless incorporate such effects into the FoM.\\
Up to this point, all three NV centers from table~\ref{tab:table1} were investigated. As the improvements are of the same order of magnitude, only NV3 is considered in the following without loss of generality. The Ramsey sensing method, which is addressed next, fulfills a similar role to the pulsed ODMR sequence and offers better sensitivities towards external DC magnetic fields~\cite{Jelezko2006,Childress2006,Barry2019} (see appendix~\ref{appd}).
\begin{figure}
    \centering
    \includegraphics[width = 0.48\textwidth]{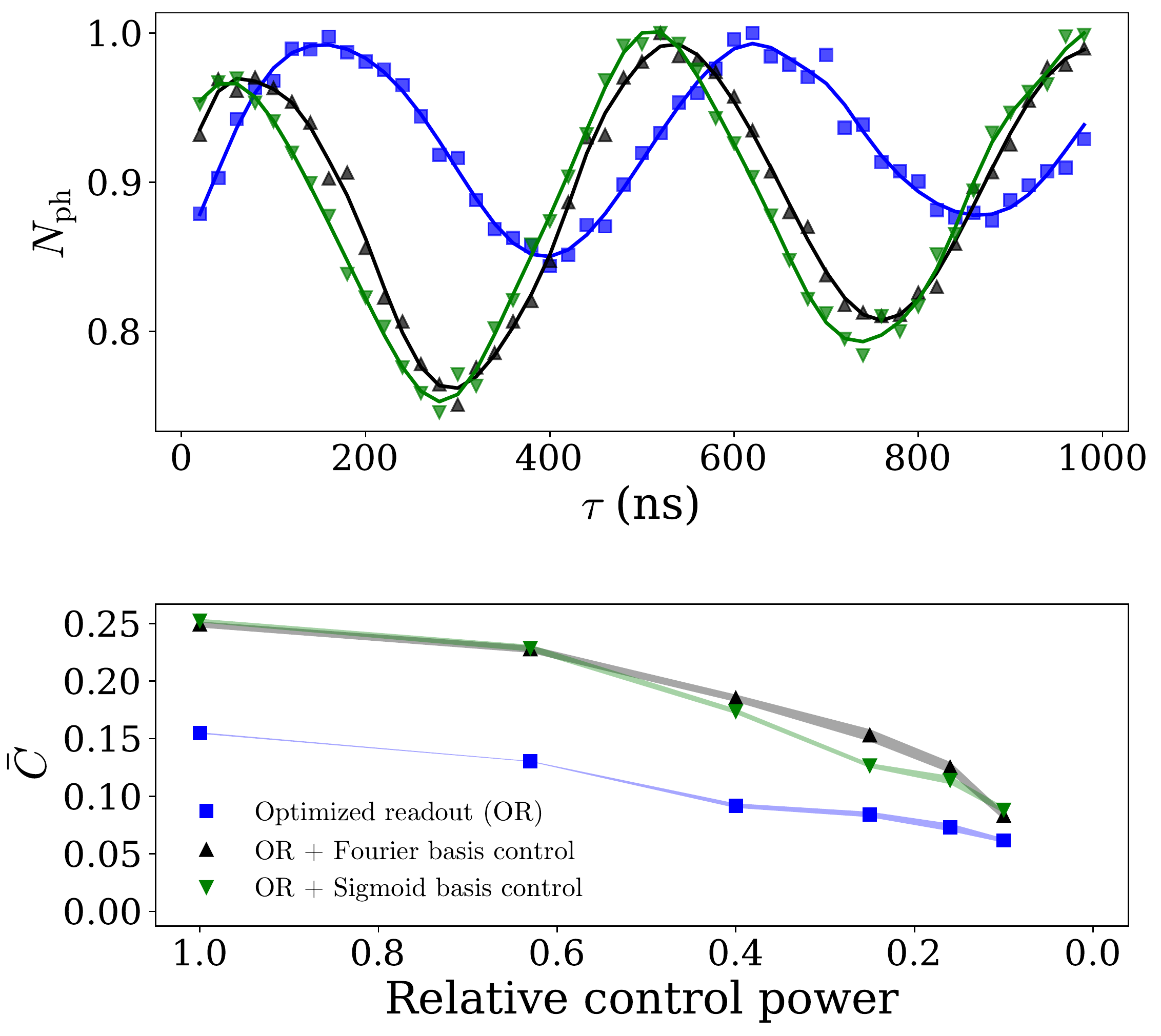}
    \caption{\label{fig:10} Optimized Ramsey measurements. The top plot shows the measurements performed at peak drive power with rectangular control pulses (blue) as well as optimized pulses in the Fourier basis (black) and Sigmoid basis (green). The optimized pulses exhibit almost double the contrast in comparison to the rectangular control pulse, with a similar precession frequency corresponding to the hyperfine levels of the transition. The length of the rectangular $\frac{\pi}{2}$-pulse is determined by performing Rabi measurements, and in this case is 67\,ns. The bottom plot shows the variation in readout contrast with respect to the change in relative control power of the control pulse. The performance of the robust optimized pulses surpasses the rectangular control pulse over the entire range of tested control power (90\% variation).}
\end{figure}
\subsection{\label{sec:4.C}Ramsey Measurement}
The Ramsey method is a type of interference measurement for DC magnetic fields. As discussed in section~\ref{sec:2} it consists of two $\frac{\pi}{2}$-pulses and offers a higher sensitivity in comparison to the ODMR methods. It should be noted that previous optimizations for D-Ramsey pulse sequences with NV centers were performed in an open-loop scheme using a cooperative design~\cite{Konzelmann2018}. Our results are obtained through a closed-loop optimization and directly quantified on the setup. The $\frac{\pi}{2}$-pulses are optimized via assessment of the contrast for a range of drive amplitudes (see Eq.~\eqref{eq:11}) via the bandwidth-limited approach discussed in section~\ref{sec:3.B} (see Fig.~\ref{fig:5}). The initial guess resembles a standard rectangular $\frac{\pi}{2}$-pulse, whose length is determined by observing Rabi Oscillations. The resulting interference fringes are shown in Fig.~\ref{fig:10}. This optimization is carried out in presence of a bias external magnetic field ($\text{B}_{\text{NV}}=12$\,mT) and on-resonance with the $m_s = 0 \leftrightarrow \text{-}1$ transition. The fringe visibility is enhanced from 0.15 to ca. 0.24 with the Fourier basis pulse, and to ca. 0.25 with the Sigmoid basis pulse using the maximum control amplitude (pulse duration of 100\,ns). The fringe visibility is directly related to the readout contrast. An improvement in the readout contrast leads to a proportional improvement in the  sensitivity of the sensor (see Eq.~\eqref{eq:C7}). 
\begin{figure}
    \centering
    \includegraphics[width = 0.48\textwidth]{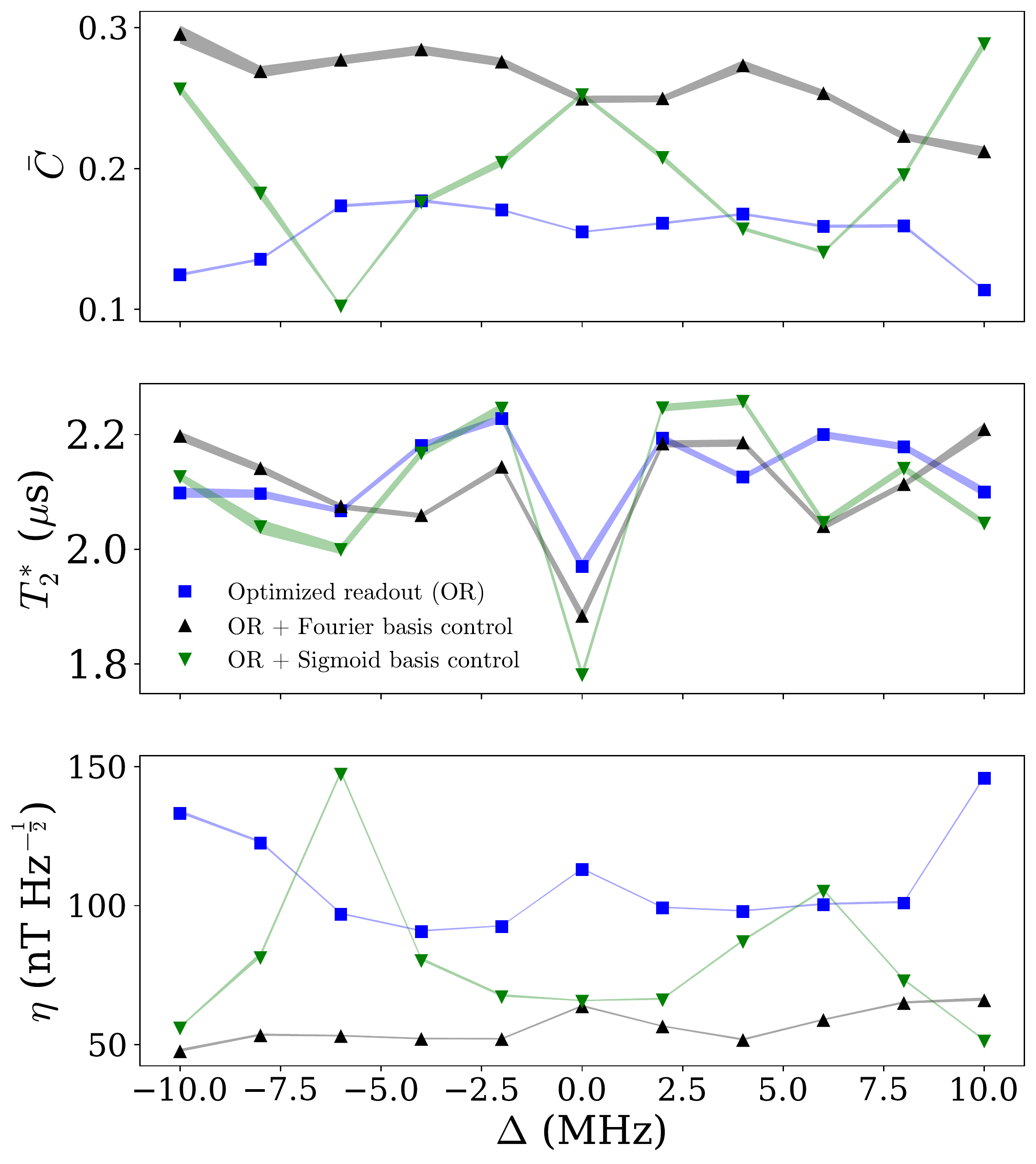}
    \caption{\label{fig:11} Comparison of Ramsey sequences with standard and optimized MW $\frac{\pi}{2}$-pulses. All measurements are performed with optimized laser pulses for the readout and the optimizations were carried out with the amplitude-robust FoM from Eq.~\eqref{eq:11}. The performance of the pulses optimized with the Fourier (black) and Sigmoid (green) basis is compared to the rectangular control pulse (blue) over a range of drive detunings $\Delta$. This range is equivalent to a variation of 0.35\,mT in the external magnetic field. The upper panel shows the readout contrast $\bar{C}$. The respective $T_2^\ast$ values are displayed in the middle panel. The bottom panel shows the resulting $\eta$. The sensitivity calculation is discussed in appendix~\ref{appd}. }
\end{figure}

The performance of the optimized pulses is further tested by performing Ramsey measurements with different drive frequencies in the vicinity of the spin transition frequency. These detunings correspond to a range of fields that could be measured in a sensing setup. The resulting readout signal summed over repeated iterations of the experiment is assessed for average sensitivity~\cite{Taylor2008} $\eta$ of the NV center. 
The sensitivity of the Ramsey sequence depends on the readout contrast and the dephasing time during the measurements (see appendix~\ref{appd}). Figure~\ref{fig:11} shows the readout contrasts, the $T_2^\ast$-times, and the average sensitivities obtained by a series of Ramsey measurements.
The Fourier pulse displays a constant readout contrast in the frequency range of $\pm 10\,$MHz. This range in the frequency corresponds to around $\pm 0.35\,$mT of variation in $B_\text{NV}$. In comparison, the Sigmoid pulse shows a marginally better $\bar{C}$ around the resonance frequency but varies strongly for different detunings. The frequency components of the pulse depend on the pulse shape (see appendix~\ref{appb}). Here, the spectrum of the Sigmoid pulse contains minima at a detuning of approximately $\pm5\,$MHz (Fig.~\ref{fig:11}). From a control perspective, such frequency selective applications are attractive for spectral hole burning~\cite{atomic2005} and quantum logic gates for superconducting qubits~\cite{Theis_2018}. The $T_2^\ast$-time is comparatively lower on resonance than off resonance for all pulses. This is due to the destructive interference of the hyperfine transition associated with the spin resonance~\cite{ladd2010}. Readout contrast enhancement inherently involves strong contributions from all the hyperfine transitions, resulting in a trade-off between $\bar{C}$ and $T_2^\ast$.
The measurements with the Fourier pulse exhibit a robust $\eta$ of less than 65\,nT\,Hz$^{-\frac{1}{2}}$. These levels of sensitivities are on par with the ones reported for single NV-based diamond scanning probes~\cite{Radtke2019,Sun2021,Grinolds2013}.

The method from this section could be generalized to replace the spin-refocusing $\pi$-pulse in other sensing methods. This would require applying the optimized $\frac{\pi}{2}$-pulses twice, using a strategy similar to the one discussed in section~\ref{sec:3.B}. Such refocusing pulses form the main building block for AC magnetic field sensing~\cite{Rembold2020}. 
\section{\label{sec:5}Conclusion}
The optimizations in this work focused on three essential parts of quantum sensing with NV centers: optical spin state readout, population inversion, and $\frac{\pi}{2}$-pulses. All three were improved for sensing methods with single NV centers, considering control power variations of up to 90\%.
Such robustness enables the sensing of larger microstructures by increasing the explorable sample area and makes the pulses more robust against experimental drift over time.
The resulting protocols are realized by replacing the building blocks of common laser and MW based schemes with optimized equivalents. The optimizations are based on a set of figures of merit which are directly measurable via contrast using a varying MW power. The feedback-based approach inherently takes experimental imperfections and unknown system parameters into account. Initially, we optimized the optical readout/initialization process, improving the spin readout contrast by 32\% in comparison to the standard protocol. Moreover, additionally optimizing the spin inversion pulse in a pulsed ODMR protocol allowed for an overall contrast improvement by 36\% leading to sub-$\mu$T\,Hz$^{-\frac{1}{2}}$ sensitivity that is maintained over a large range of MW amplitudes. Such robust excitation pulses lead to a large interrogation volume. Especially, for ensembles of NV centers this results in improved readout counts for a larger area, and in turn, enhanced sensitivity~\cite{Barry2019}. 
To maintain frequency sensitivity, different optimization bases were explored. The Sigmoid basis leads to spin transfer within a limited bandwidth envelope, reducing the off-resonant excitation. Additionally, we obtained an optimized $\frac{\pi}{2}$-pulse for Ramsey measurements, enhancing the fringe contrast by 67\% with respect to the square pulse with pre-optimized optical readout at maximum control power. Consequently, we obtained a two-fold enhancement in the average sensitivities, ranging below 100\,nT\,Hz$^{-\frac{1}{2}}$ over a set of induced bias field strengths. While we applied the optimization to shallow NV centers, the approach is straightforwardly applicable to other NV-based systems like diamond scanning probes and NV ensembles used for wide-field imaging where similar control robustness features are required.\\
\begin{acknowledgments}
We thank A. Marshall, F. Motzoi, R. Nelz, S. Z. Ahmed, T. Reisser and M. Rossignolo for their insights and suggestions. We furthermore thank T. Reisser for his assistance with the RedCRAB software suite. This work has received funding from the European Union's Horizon 2020 research and innovation program under the Marie Sk\l{}odowska-Curie grant agreement N$^\circ$ 765267 (QuSCo) and under the grant agreement N$^\circ$ 820394 (ASTERIQS).

\textbf{Authors' contributions}: NO and EN planned the experiments. PR, NO, and MM planned the optimization strategies and worked to facilitate the remote optimization. NO designed and performed the experiments and analyzed the data. TC, SM, EN and MM supervised the project. All the authors discussed the results and contributed to the manuscript.
\end{acknowledgments}

\bibliography{apssamp.bib}

\appendix

\section{\label{appa}Experimental setup}

All the measurements were carried out on a custom-built confocal setup, with excitation wavelength of 520 nm (Swabian instruments, DL nSec, PE 520) and objective numerical aperture of 0.8 (Olympus, LMPLFLN100X). Rejection of the out-of-focus fluorescence signal was achieved by using single mode optical fibers (Thorlabs, SM450 and SM600) at the excitation and detection arm of the confocal microscope. Further, contributions from the $NV^0$ charge state were blocked with a spectral filter (Thorlabs, FEL0600, Longpass 600 nm) in the detection arm. Fluorescence signal from the single NVs was detected with a single photon counting module (APD, Excelitas, SPCM-AQRH-14, quantum efficiency $\approx$68\%) and the acquired data was logged with a data acquisition card (National Instruments, PCIe-6323). Second order intensity correlation measurements were performed with a Hanbury-Brown Twiss setup attached to a time resolved counting device (PicoQuant, PicoHarp 300). The fluorescence signal was filtered and analyzed with a spectrometer (SP-2500, Princeton Instruments) to ensure the charge state stability in the diamond sample. The MW control pulses were generated with IQ mixing with the MW signal generator source (Tektronix 4104A, IF bandwidth of 400 MHz). The in-phase and quadrature components were obtained with an arbitrary waveform generator (AWG, Tabor 1204 A, 2.3 GSa s$^{-1}$). Control pulses were delivered to the diamond sample with a custom-built $\Omega$-shaped antenna~\cite{Opaluch2021} after amplification (ZHL-16W-43-S+, Mini-Circuits, typ. +45 dB). Channel synchronization was ensured using a sync device (Swabian Instruments, Pulse Streamer 8/2) to trigger the diode laser, AWG, MW source, APD count window and the data acquisition device. The sample along with the MW-antenna was mounted onto a piezo-scanner (Physik Instrumente (PI), P-611.3O) to perform the confocal scans and address individual NV centers. The remote connection to the optimization server was obtained via a combination of MATLAB (remote system) and Python (RedCRAB GUI) based control programs.

\section{\label{appb}Random bases for dCRAB optimization}

In the dCRAB algorithm, random bases are used whose elements can be defined through a superparameter $\omega$ which stays constant throughout the optimization. In this work, we have used two different bases, referred to as the Fourier and the Sigmoid basis. They differ in their shape and properties.

The Fourier basis is most commonly used with dCRAB. It consists of $M=2$ out of phase trigonometric elements with frequency $0\le\omega\le \omega_\text{max}$:
\begin{equation}
    \begin{split}
        f_\text{Fourier}^1(\omega; t) =& \sin(\omega t)\\
        f_\text{Fourier}^2(\omega; t) =& \cos(\omega t).
    \end{split}
\end{equation}
The Sigmoid basis~\cite{Rembold2021} consists of sigmoid functions ($M=1$) with an offset of $\epsilon \sigma \le \omega \le t_p-\epsilon \sigma$. $\epsilon$ represents an offset factor. The basis always includes one element at $\omega=\epsilon \sigma$ which is optimized in every superiteration to ensure the pulse length is constant (i.e. $u_i(t=0)=u_i(t=t_p)=0$). For the same reason, an element is added automatically with $\omega=t_p-\epsilon \sigma$ and amplitude $A=\sum_n^{N}A_n$.
\begin{equation}
    f_\text{Sigmoid}(\omega; t) = \frac{1}{\sqrt{2\pi }\sigma }\int_{0}^{t} e^{-\frac{1}{2}\left( \frac{\tau-\omega}{\sigma }\right) ^{2}}d\tau.
\end{equation}
They both have different properties. In general, the Fourier basis is bandwidth-limited through the upper limit for $\omega$. The Sigmoid basis is bandwidth-limited due to the limited rise time defined by $\sigma$.  However, in both cases higher frequency terms may be introduced through cut-offs (i.e., cut-offs in the time domain or amplitude domain).\\
It should also be noted that the basis choice determines which shapes are complex, and which are simple to produce. While the Fourier basis produces oscillations with few basis elements, the Sigmoid basis produces approximately square pulses, without cut-offs.

\section{\label{appd}Sensitivity Calculation}

NV center based sensing is fundamentally limited by the spin projection limit~\cite{budker2020}. This limit can be expressed as
\begin{equation}
    \eta_{sp} = \frac{\hbar}{Sg_{e}\mu_{B}}\frac{1}{\sqrt{t_{m}}},
\end{equation}
where, $\hbar$ is the reduced Plank's constant, $g_{e}$ is the Land\'e factor, $\mu_{B}$ is the Bohr magneton, and $t_m$ is the measurement time. In addition, optical readout processors are subjected to photon shot noise that further adheres the sensitivity. For the averaged readout process discussed in section \ref{sec:1}, Eq.~\eqref{eq:1}, the readout fidelity can be equivalently written as
\begin{equation}
    \mathcal{F} = \sqrt{1+\frac{1}{\bar{C}^2\bar{R}}}.
\end{equation}
$\bar{C}$ is the average readout contrast between the two spin states of the system and $\bar{R}$ is the average count rate. Further, an overhead cost is always involved in an experimental scenario. Really long spin initialization and readout duration deteriorates the overall sensitivity of the sensor, this can be expressed as a scaling factor for the sensitivity
\begin{equation}
    \kappa_{exp} = \sqrt{\frac{t_{m}+2\times t_{i}}{t_{m}}},
\end{equation}
under the assumption that the initialization and readout duration are equal ($t_i$). Finally, for DC magnetometry methods, the dehpasing time $T_2^\ast$ further limits the sensitivity, this can be expressed the decoherence function of the $T_2^\ast$-limited processes,
\begin{equation}
\label{eq:D4}
    f_{d} = e^{\left( \frac{t_{m}}{T_{2}^{\ast }}\right) ^{m}},
\end{equation}
where, $m$ is the order of decoherence. For spectroscopic measurement around the NV resonance peaks, the sensitivity depends on the resonance profile itself~\cite{Barry2019}. In case of Gaussian resonance profiles for the pulsed ODMR measurements, the overall sensitivity can be computed as
\begin{equation}
\label{eq:D5}
    \eta^\text{po}=\mathscr{P}\frac{1}{\gamma_\text{nv}}\frac{\sigma_f}{\bar{C}\sqrt{\bar{R}}}{\sqrt{T_{\pi}+t_{m}}}.
\end{equation}
Here, $\sigma_f$ is the resonance peak FWHM, $\gamma_\text{nv} = \frac{g_{e}\mu_{B}}{\hbar}$, is the gyromagnetic ratio of the NV spin, and  $T_\pi$ is the pulse duration. The factor $\mathscr{P}$ relates to the shape of the resonance, for a Gaussian profile $\mathscr{P}$ = $\sqrt{\frac{e}{8\text{ln}2}}$~\cite{Dreau2011}. For shorthand notation, the measurement time is assumed to involve the overhead experimental time $t_m = t_{w} + 2\times t_i$. The relevant parameters for the sensitivity calculation in section~\ref{sec:4} were obtained by fitting the normalized count with the following function:
\begin{equation}
\label{eq:c6}
    N_\text{ph}^\text{po}(f)=\bar{R}\times \left[1-\bar{C}\times e^{-\frac{1}{2}\left( \frac{f-f_{0}}{\Delta f}\right) ^{2} }\right],
\end{equation}
where $f_0$ is the resonance peak, the normalized counts are calculated by dividing the data with the baseline counts (counts away from the resonance, where no spin transfer occurs). For a Gaussian profile, $\sigma_f = 2\sqrt{2\ln2}\times \Delta f$. It is noteworthy that for pulsed ODMR measurements at low MW power, $T_2^\ast$-limit becomes relevant and has to be considered for sensitivity calculations, the reader is advised to refer to Ref.~\cite{Dreau2011} for more details. 

The average sensitivity for the Ramsey sequence based methods can be expressed under the $T_2^\ast$-limit as,
\begin{equation}
\label{eq:C7}
    \eta ^\text{Ra}=\frac{1}{\bar{C}\gamma_\text{nv}\tau} \mathrm{exp}\left[\left( \frac{\tau }{T_{2}^{\ast }}\right) ^{m}\right] \sqrt{\tau +t_{m}}.
\end{equation}
The free induction decay of the Ramsey fringes for single NV centers highlight the hyperfine structure originating from the electron-nuclear spin coupling. Likewise, the related normalized readout counts can be fitted with a sum of the three precessing hyperfine transitions,
\begin{equation}
\scalemath{0.75}{
    N_\text{ph}^\text{Ra}(t)=\bar{R}\left[ 1+\left(\bar{C}\times e^{-(\tau/T^{\ast}_2)^{m}}\sum ^{3}_{i}A_{i}\cos \left( 2\pi \nu_{i}t+\phi_{i}\right)\right)\right]},
\end{equation}
where, $\nu_{i}$ and $\phi_{i}$ are the precession frequency and phase corresponding to the hyperfine transitions. The values for $\nu_{i}$ depend on the detuning of the drive with respect to the transitions in an external bias field (see Fig.~\ref{fig:11}). The sensitivities in Fig.~\ref{fig:11} are obtained at $\tau$ = 0.5$\times T^{\ast}_2$. The normalized readout count in this case is obtained by dividing the data with the $m_s = 0$ readout count.

\end{document}